\documentclass[useAMS,usenatbib]{mnras}

\usepackage{graphicx, bm, amssymb, xcolor}
\usepackage{verbatim}
\usepackage[all]{hypcap}
\usepackage{xspace}
\usepackage{multirow,bigdelim}
\usepackage[fleqn]{amsmath}

\usepackage{epsfig}
\usepackage{aas_macros}
\usepackage{natbib}
\usepackage{times,txfonts}
\usepackage{morefloats}
\usepackage{pdflscape}

\newcommand{\msun}{\mbox{M$_\odot$}}
\newcommand{\yr}{\mbox{${\rm yr}$}}

\newcommand{\gyr}{\mbox{${\rm Gyr}$}}
\newcommand{\pc}{\mbox{${\rm pc}$}}
\newcommand{\mpc}{\mbox{${\rm Mpc}$}}
\newcommand{\kpc}{\mbox{${\rm kpc}$}}
\newcommand{\kms}{\mbox{${\rm km}~{\rm s}^{-1}$}}
\newcommand{\cmc}{\mbox{${\rm cm}^{-3}$}}

\newcommand{\dex}{\mbox{${\rm dex}$}}
\newcommand{\feh}{\mbox{$[{\rm Fe}/{\rm H}]$}}
\newcommand{\afe}{\mbox{$[\alpha/{\rm Fe}]$}}
\newcommand{\iqr}{\mbox{${\rm IQR}$}}
\newcommand{\riqr}{\mbox{$r_{\rm IQR}$}}
\newcommand{\dfehdt}{\mbox{${\rm d}\feh/{\rm d}\log{t}$}}
\newcommand{\ngc}{\mbox{$N_{\rm GC}$}}
\newcommand{\ngcp}{\mbox{$N_{\rm GC}'$}}
\newcommand{\sfr}{\mbox{${\rm SFR}$}}

\newcommand{\mvir}{\mbox{$M_{200}$}}

\newcommand{\vmax}{\mbox{$V_{\rm max}$}}

\newcommand{\cnfw}{\mbox{$c_{\rm NFW}$}}
\newcommand{\ttf}{\mbox{$\tau_{25}$}}
\newcommand{\tfz}{\mbox{$\tau_{50}$}}

\newcommand{\za}{\mbox{$z_{\rm a}$}}
\newcommand{\tf}{\mbox{$\tau_{\rm f}$}}

\newcommand{\thub}{\mbox{$\tau_{\rm H}$}}
\newcommand{\nbrz}{\mbox{$N_{{\rm br},z>2}$}}
\newcommand{\nbr}{\mbox{$N_{\rm br}$}}
\newcommand{\rz}{\mbox{$r_{z>2}$}}
\newcommand{\nleaf}{\mbox{$N_{\rm leaf}$}}

\newcommand{\fcorr}{\mbox{$f_{\rm corr}$}}
\newcommand{\mosaics}{MOSAICS\xspace}
\newcommand{\emosaics}{E-MOSAICS\xspace}
\newcommand{\eagle}{EAGLE\xspace}

\newcommand{\be}{\begin{equation}}
\newcommand{\ee}{\end{equation}}
\newcommand{\bea}{\begin{eqnarray}}
\newcommand{\eea}{\end{eqnarray}}

\title[Formation and assembly history of the Milky Way]{\vspace{-7mm}The formation and assembly history of the Milky Way revealed by its globular cluster population\vspace{-5mm}}
\author{J.~M.~Diederik Kruijssen,$^{1}$\thanks{E-mail: \href{kruijssen@uni-heidelberg.de}{kruijssen@uni-heidelberg.de}} Joel~L.~Pfeffer,$^2$ Marta Reina-Campos,$^1$
\newauthor Robert~A.~Crain$^2$ and Nate Bastian$^2$ \\
$^1$Astronomisches Rechen-Institut, Zentrum f\"{u}r Astronomie der Universit\"{a}t Heidelberg, M\"{o}nchhofstra\ss e 12-14, 69120 Heidelberg, Germany\\
$^2$Astrophysics Research Institute, Liverpool John Moores University, IC2, Liverpool Science Park, 146 Brownlow Hill,
Liverpool L3 5RF, United Kingdom\vspace{-4mm}}

\begin{document}

\date{Accepted 2018 June 12. Received 2018 June 8; in original form 2018 March 25\vspace{-3mm}}

\pagerange{\pageref{firstpage}--\pageref{lastpage}} \pubyear{2018}

\maketitle

\label{firstpage}

\begin{abstract}
We use the age-metallicity distribution of 96 Galactic globular clusters (GCs) to infer the formation and assembly history of the Milky Way (MW), culminating in the reconstruction of its merger tree. Based on a quantitative comparison of the Galactic GC population to the 25 cosmological zoom-in simulations of MW-mass galaxies in the \emosaics project, which self-consistently model the formation and evolution of GC populations in a cosmological context, we find that the MW assembled quickly for its mass, reaching $\{25,50\}\%$ of its present-day halo mass already at $z=\{3,1.5\}$ and half of its present-day stellar mass at $z=1.2$. We reconstruct the MW's merger tree from its GC age-metallicity distribution, inferring the number of mergers as a function of mass ratio and redshift. These statistics place the MW's assembly {\it rate} among the 72th-94th percentile of the \emosaics galaxies, whereas its {\it integrated} properties (e.g.~number of mergers, halo concentration) match the median of the simulations. We conclude that the MW has experienced no major mergers (mass ratios $>$1:4) since $z\sim4$, sharpening previous limits of $z\sim2$. We identify three massive satellite progenitors and constrain their mass growth and enrichment histories. Two are proposed to correspond to Sagittarius (few~$10^8~\msun$) and the GCs formerly associated with Canis Major ($\sim10^9~\msun$). The third satellite has no known associated relic and was likely accreted between $z=0.6$--$1.3$. We name this enigmatic galaxy {\it Kraken} and propose that it is the most massive satellite ($M_*\sim2\times10^9~\msun$) ever accreted by the MW. We predict that $\sim40\%$ of the Galactic GCs formed ex-situ (in galaxies with masses $M_*=2\times10^7$--$2\times10^9~\msun$), with $6\pm1$ being former nuclear clusters.\looseness=-1
\end{abstract}

\begin{keywords}
Galaxy: evolution --- Galaxy: formation --- (Galaxy:) globular clusters: general --- Galaxy: halo --- Galaxy: stellar content --- galaxies: star formation\vspace{-6mm}
\end{keywords}

\section{Introduction} \label{sec:intro}

It has been known for more than half a century that the Milky Way formed through a process of gravitational collapse \citep{eggen62}. In the decades since, it has become clear that galaxy formation is a continuous hierarchical process, in which central hubs grow through a combination of rapid, in-situ star formation (dominant at early cosmic times) and the gradual accretion of smaller satellites \citep[in the context of this work, see e.g.][]{white78,searle78,white91,forbes97,cote98,cote00,oser10,beasley18}. More recently, deep observations of stellar haloes are revealing the relics of this ongoing accretion process, during which dwarf galaxies are tidally disrupted and their debris builds up the central galaxy's halo \citep[e.g.][]{ibata94,martin04,belokurov06,mcconnachie09,grillmair09,crnojevic16}.

While the relics of the galaxy formation and assembly process are now found in haloes throughout the local Universe, a comprehensive picture of the formation and assembly history of the Milky Way is still lacking. The existing constraints on the Galaxy's assembly history are generally based on its most recent accretion events \citep[e.g.][]{martin04,belokurov06} or on indirect inference using comparisons to (cosmological) simulations of galaxy formation and evolution \citep[e.g.][Hughes et al.~in prep.]{wyse01,stewart08,shen10,deason13,mackereth18}. These approaches have provided two major insights. Firstly, the hierarchical growth of the Milky Way through satellite accretion is still continuing at the present day. Secondly, these accretion events predominantly correspond to minor mergers, with the last major merger having taken place at $z>2$. An extensive body of literature aims to refine these results and obtain a complete understanding of the Galaxy's formation and assembly history (see e.g.~\citealt{belokurov13} for a recent review).

In this paper, we constrain the formation and assembly history of the Milky Way by applying a new technique that combines and expands on previous approaches -- we perform a quantitative and detailed comparison to a suite of cosmological zoom-in simulations and use the insights from this comparison to reconstruct the merger tree of the Milky Way. Specifically, we draw inspiration from the seminal paper by \citet{searle78}, who used the Galactic globular cluster (GC) population to infer that the GCs in the Galactic halo formed over a longer timespan than the GCs associated with the Galactic bulge, leading to the conclusion that the Milky Way grew by hierarchical galaxy growth. This idea has since been refined -- improved measurements of the ages of GCs have shown that the distribution of Galactic GCs in age-metallicity space is bifurcated \citep[e.g.][]{marinfranch09}, with a young and metal-poor branch tracing part of the GC population thought to result from satellite galaxy accretion \citep[e.g.][]{forbes10,leaman13}. These results suggest that the detailed age-metallicity distribution of GCs may be used to infer the accretion history of the host galaxy.

The key ingredient enabling us to extract the wide range of information encoded in the Galactic GC population is the \emosaics\footnote{This is an acronym for `MOdelling Star cluster population Assembly In Cosmological Simulations within EAGLE'.} project \citep{pfeffer18,kruijssen18b}. Cluster formation and disruption are well known to depend on the local environmental conditions within galaxies \citep[see recent reviews by][]{kruijssen14c,forbes18}. Modelling the co-formation and co-evolution of GC populations and their host galaxies thus requires following the spatial and kinematic phase space distributions of the cluster populations self-consistently in their cosmological context. This prohibits the use of models based on semi-analytic or particle tagging techniques. Instead, \emosaics self-consistently couples environmentally-dependent, sub-grid models for the formation, evolution, and tidal disruption of stellar clusters \citep[MOSAICS,][]{kruijssen11,kruijssen12c,pfeffer18} to the locally-resolved baryonic conditions obtained with the state-of-the-art cosmological galaxy formation model \eagle \citep{schaye15,crain15}. The initial set of 25 hydrodynamical cosmological zoom-in simulations of Milky Way-mass galaxies allows us to follow the co-formation and co-evolution of galaxies and their GC populations.

Specifically, we show in \citet{kruijssen18b} that the distribution of GCs in age-metallicity space is a sensitive probe of the host galaxy's formation and assembly history. The GC age-metallicity distributions obtained in \emosaics exhibit strong galaxy-to-galaxy variations, resulting from differences in their evolutionary histories. By correlating several metrics describing the GC age-metallicity distributions of the 25 simulated galaxies with quantities describing their formation and assembly histories (such as halo properties, formation and assembly redshifts, stellar mass assembly time-scales, numbers of major and minor mergers, redshift distribution of mergers), we identify a set of 20 correlations with high statistical significance. These correlations show that the GC age-metallicity distribution is a sensitive probe of the host galaxy's stellar mass growth, metal enrichment, and minor merger history. Moreover, it is shown that the mass dependence of galaxy evolution in age-metallicity space enables the use of individual GCs to infer the masses and metallicities of their natal galaxies as a function of redshift. We demonstrate that this can be used to reconstruct the merger tree of the host galaxy. Applying any of these individual results to the GC population of the Milky Way may already significantly increase our understanding of its formation and assembly history. In this paper, we take the comprehensive approach of employing the full range of analysis tools presented in \citet{kruijssen18b} to infer how the Galaxy formed and assembled.

This method represents a critical first step in our ongoing efforts to unlock the potential of GCs as quantitative tracers of galaxy formation and assembly. However, its applicability is presently restricted to a handful of galaxies due to observational limitations. Even with current state-of-the-art observatories and stellar population synthesis models, the main obstacle in applying our framework is the availability of reliable GC ages. The precision of GC age measurements is a strong function of distance due to the different methods involved, with uncertainties of $\sigma(\tau)=1$--$2~\gyr$ in the Milky Way \citep[$D\sim10~\kpc$,][]{dotter10,dotter11,vandenberg13}, $\sigma(\tau)\sim2~\gyr$ in the Large Magellanic Cloud \citep[LMC; $D\sim50~\kpc$,][]{wagnerkaiser17b}, and $\sigma(\tau)\sim4~\gyr$ in the dwarf galaxy NGC4449 \citep[$D\sim3.8~\mpc$,][]{annibali18}. For the foreseeable future, the Milky Way, the LMC, and possibly M31 are thus the most suitable targets for applying our technique.\footnote{This adds to the growing body of literature showing that decreasing the uncertainties on GC ages from resolved or integrated-light spectroscopy is one of the most pressing open problems in GC research (also see \citealt{forbes18} for a recent discussion).}

For galaxies with reliable GC age measurements, the potential of using the GC age-metallicity distribution for constraining galaxy formation and assembly histories is clear. In this work, we therefore apply our framework to the GC population of the Milky Way. As will be shown below, this provides important new insights into the formation and assembly history of our host galaxy.

\section{Cosmological zoom-in simulations of GC populations and their host galaxies} \label{sec:model}

In this work, we interpret the age-metallicity distribution of Galactic GCs in terms of the Milky Way's formation and assembly history using the \emosaics simulations of Milky Way-mass galaxies and their GC populations \citep{pfeffer18,kruijssen18b}. These simulations combine the \mosaics model for cluster formation and evolution \citep{kruijssen11,pfeffer18} with the \eagle model for galaxy formation simulations \citep{schaye15,crain15}. All simulations adopt a $\Lambda$CDM cosmogony, described by the parameters advocated by the \citet{planck14}, namely $\Omega_0 = 0.307$, $\Omega_{\rm b} = 0.04825$, $\Omega_\Lambda= 0.693$, $\sigma_8 = 0.8288$, $n_{\rm s} = 0.9611$, $h = 0.6777$, and $Y = 0.248$. Before proceeding, we briefly summarise the physical ingredients contained in these models.

\subsection{Summary of the models}
\eagle uses a modified version of the $N$-body TreePM smoothed particle hydrodynamics code \textsc{Gadget~3} \citep[last described by][]{springel05c} and includes a wide array of subgrid routines for modelling the baryonic physics governing galaxy formation and evolution. Most importantly, these include radiative cooling and photoionization heating \citep{wiersma09}, a redshift-dependent UV background \citep{haardt01}, stochastic star formation within gas with density above a metallicity-dependent threshold \citep{schaye04,schaye08}, stellar feedback \citep{dallavecchia12}, metal enrichment \citep{wiersma09b}, the growth of supermassive black holes by gas accretion and merging \citep{springel05b,rosasguevara15,schaye15}, and feedback from active galactic nuclei \citep{booth09,schaye15}. The application of this galaxy formation model in \eagle broadly reproduces the evolution of galaxy stellar masses and sizes \citep{furlong15,furlong17}, their photometric properties \citep{trayford15} and cold gas content \citep[e.g.][]{lagos15,bahe16,crain17}, and key properties of the circumgalactic medium \citep[e.g.][]{rahmati15,oppenheimer16}.

In \emosaics, the \eagle galaxy formation model is combined with the \mosaics cluster model. Stellar clusters are formed as a subgrid component of new born stellar particles based on the local gas properties at the time of their formation. The fraction of star formation occurring in bound clusters \citep[the cluster formation efficiency]{bastian08,goddard10} is environmentally dependent according to the model of \citet{kruijssen12d}. The initial cluster mass function follows a power law with a universal slope of $-2$ between $M=10^2~\msun$ and an environmentally dependent upper mass scale from \citet{reinacampos17}. Both the cluster formation efficiency and the upper cluster mass scale increase with the gas pressure, implying that actively star-forming environments such as high-redshift galaxies or local-Universe galaxy mergers form a larger fraction of their mass in clusters and also form more massive ones than in low-pressure environments \citep[see recent reviews by][]{kruijssen14c,forbes18}. This cluster formation model has been shown to reproduce a wide range of observations \citep[e.g.][]{adamo15b,johnson16,freeman17,ward18,messa18}.

The clusters evolve in an environmentally-dependent fashion according to four disruption mechanisms. Firstly, stellar clusters can lose up to 40~per~cent of their mass by stellar evolution, for which we adopt the model natively used in \eagle. Secondly, we model the mass loss due to tidal shocks from the interstellar medium (ISM) by calculating the tidal tensor on the fly at the position of each cluster \citep{kruijssen11}. \emosaics thus self-consistently captures the disruptive power of the different galactic environments experienced by a cluster throughout its lifetime, which provides a much improved description of tidal disruption relative to previous work. However, the absence of a cold ISM in \eagle causes cluster disruption by tidal shocks to be underestimated in environments with low and intermediate gas pressures ($P/k<10^7~{\rm K}~\cmc$, see \citealt{pfeffer18}). Tidal shocks are still the dominant disruption mechanism in high-pressure environments. Thirdly, we include mass loss by two-body interactions between the stars in a cluster \citep[see also \citealt{lamers05b}]{kruijssen11}. This mechanism is relevant in low-density environments, where the ISM is not as disruptive. Finally, we include a post-processing model for the destruction of the most massive clusters due to dynamical friction \citep{pfeffer18}. This cluster disruption model has been shown to reproduce observed cluster age distributions, mass distributions, spatial distributions, and kinematics \citep{kruijssen11,kruijssen12c,miholics17,adamo18}.

\subsection{Simulated galaxy properties and suitability for comparison to the Milky Way}
The \emosaics suite contains 25 zoom-in simulations of all Milky Way-mass galaxies in the \eagle Recal-L025N0752 volume, spanning present-day halo masses of $7\times10^{11}<\mvir/\msun<3\times10^{12}$. This is an unbiased and volume-limited galaxy sample, with no selection on any other galaxy properties. As shown by \citet{pfeffer18} and \citet{kruijssen18b}, these galaxies cover a wide range of formation and assembly histories, making it the ideal reference sample for inferring the formation and assembly history of the Milky Way from the observed properties of the Galactic GC population. We demonstrate the suitability of the \emosaics galaxies for this purpose in \autoref{tab:sims}, which shows that the key physical properties of the Milky Way are consistent with those spanned by the simulated galaxies. Specifically, we show the halo mass, stellar mass, star-forming gas mass (cf.~atomic and molecular gas), non-star-forming gas mass (cf.~hot halo gas), current star formation rate (SFR), stellar bulge mass, bulge-to-total stellar mass ratio, stellar disc mass (cf.~the combined thin and thick stellar disc), and the stellar disc scale radius. To isolate the stellar disc of each galaxy, we select all stellar particles with specific angular momentum $J_z/J_{\rm circ}>0.5$ \citep[cf.~fig.~2 of][]{abadi03b}.

\begin{table*}
  \caption{Properties of the 25~Milky Way-mass, $L^*$ galaxies at $z=0$ in the cosmological zoom-in simulations considered in this work. From left to right, the columns show: simulation ID; log halo mass; log stellar mass; log star-forming gas mass; log non-star-forming gas mass; SFR averaged over the last 300~Myr; log bulge mass; bulge-to-total mass ratio; log stellar disc mass; stellar disc scale radius; log stellar halo mass. All of the integrated baryonic galaxy properties are measured within 30~proper kpc. To give an indication of the typical values and dynamic ranges, we also include four rows listing the minimum, median, and maximum values, as well as the interquartile range of each column. For reference, the final row shows the properties of the Milky Way from \citet{blandhawthorn16} including the uncertainties, where $M_{\rm d}$ includes both the thin disc and the thick disc, and $R_{\rm d}$ represents the mass-weighted average scale length of the total disc. The Milky Way's cold gas mass ($M_{\rm SF}$) is taken from \citet{nakanishi16}. This table shows that the Milky Way is consistent with the range of properties spanned by the \emosaics galaxies.}
\label{tab:sims}
  \begin{tabular}{lccccccccc}
   \hline
   Name & $\log \mvir$ & $\log M_*$ & $\log M_{\rm SF}$ & $\log M_{\rm NSF}$ & \sfr & $\log M_{\rm b}$ & $M_{\rm b}/M_*$ & $\log M_{\rm d}$ & $R_{\rm d}$ \\ 
     & $[\msun]$ & $[\msun]$ & $[\msun]$ & $[\msun]$ & $[\msun~\yr^{-1}]$ & $[\msun]$ &  & $[\msun]$ & $[\kpc]$  \\ 
   \hline
   MW00 & $11.95$ & $10.28$ & $9.39$ & $10.34$ & $0.63$ & $9.67$ & $0.24$ & $10.16$ & $6.40$ \\ 
   MW01 & $12.12$ & $10.38$ & $9.55$ & $11.05$ & $0.93$ & $10.16$ & $0.62$ & $9.96$ & $3.13$ \\ 
   MW02 & $12.29$ & $10.56$ & $9.82$ & $11.19$ & $1.65$ & $9.85$ & $0.19$ & $10.47$ & $7.36$ \\ 
   MW03 & $12.17$ & $10.42$ & $9.82$ & $11.04$ & $1.72$ & $9.83$ & $0.26$ & $10.29$ & $6.53$ \\ 
   MW04 & $12.02$ & $10.11$ & $9.29$ & $10.84$ & $0.35$ & $9.71$ & $0.39$ & $9.90$ & $5.11$ \\ 
   MW05 & $12.07$ & $10.12$ & $8.51$ & $10.32$ & $0.08$ & $9.76$ & $0.44$ & $9.87$ & $2.15$ \\ 
   MW06 & $11.96$ & $10.31$ & $9.89$ & $10.86$ & $2.44$ & $9.62$ & $0.20$ & $10.21$ & $4.91$ \\ 
   MW07 & $11.86$ & $10.16$ & $9.81$ & $10.86$ & $1.52$ & $9.67$ & $0.33$ & $9.98$ & $3.92$ \\ 
   MW08 & $11.87$ & $10.12$ & $9.34$ & $10.78$ & $1.08$ & $9.45$ & $0.21$ & $10.02$ & $2.99$ \\ 
   MW09 & $11.87$ & $10.16$ & $9.62$ & $10.52$ & $1.36$ & $9.68$ & $0.33$ & $9.98$ & $4.42$ \\ 
   MW10 & $12.36$ & $10.48$ & $9.47$ & $11.38$ & $0.93$ & $10.26$ & $0.61$ & $10.07$ & $3.85$ \\ 
   MW11 & $12.15$ & $10.06$ & $9.26$ & $11.04$ & $0.63$ & $9.90$ & $0.69$ & $9.56$ & $2.24$ \\ 
   MW12 & $12.34$ & $10.44$ & $9.69$ & $11.36$ & $1.13$ & $10.12$ & $0.47$ & $10.16$ & $4.75$ \\ 
   MW13 & $12.38$ & $10.37$ & $8.81$ & $11.22$ & $0.49$ & $9.85$ & $0.30$ & $10.21$ & $2.72$ \\ 
   MW14 & $12.34$ & $10.59$ & $9.70$ & $11.45$ & $1.91$ & $10.23$ & $0.44$ & $10.34$ & $3.85$ \\ 
   MW15 & $12.16$ & $10.15$ & $9.78$ & $10.99$ & $2.15$ & $10.02$ & $0.74$ & $9.56$ & $3.95$ \\ 
   MW16 & $12.32$ & $10.54$ & $7.16$ & $10.40$ & $0.00$ & $10.43$ & $0.77$ & $9.90$ & $2.33$ \\ 
   MW17 & $12.29$ & $10.49$ & $9.42$ & $10.97$ & $0.57$ & $10.31$ & $0.67$ & $10.00$ & $2.76$ \\ 
   MW18 & $12.25$ & $10.00$ & $9.40$ & $10.75$ & $1.31$ & $9.80$ & $0.63$ & $9.57$ & $6.09$ \\ 
   MW19 & $12.20$ & $9.93$ & $9.64$ & $11.20$ & $1.13$ & $9.64$ & $0.52$ & $9.61$ & $6.90$ \\ 
   MW20 & $11.97$ & $10.10$ & $9.60$ & $11.04$ & $0.71$ & $9.69$ & $0.38$ & $9.89$ & $3.87$ \\ 
   MW21 & $12.12$ & $10.03$ & $9.25$ & $10.60$ & $0.50$ & $9.94$ & $0.82$ & $9.30$ & $6.20$ \\ 
   MW22 & $12.15$ & $10.43$ & $7.55$ & $10.33$ & $0.00$ & $10.33$ & $0.80$ & $9.73$ & $3.39$ \\ 
   MW23 & $12.19$ & $10.53$ & $9.97$ & $11.24$ & $3.30$ & $9.94$ & $0.26$ & $10.40$ & $6.55$ \\ 
   MW24 & $12.06$ & $10.29$ & $9.40$ & $10.74$ & $0.65$ & $9.93$ & $0.43$ & $10.05$ & $2.62$ \\ 
   \hline
   Minimum & $11.86$ & $9.93$ & $7.16$ & $10.32$ & $0.00$ & $9.45$ & $0.19$ & $9.30$ & $2.15$ \\ 
   Median & $12.15$ & $10.29$ & $9.47$ & $10.97$ & $0.93$ & $9.85$ & $0.44$ & $9.98$ & $3.92$ \\ 
   Maximum & $12.38$ & $10.59$ & $9.97$ & $11.45$ & $3.30$ & $10.43$ & $0.82$ & $10.47$ & $7.36$ \\ 
   IQR & $0.27$ & $0.32$ & $0.41$ & $0.45$ & $0.95$ & $0.43$ & $0.33$ & $0.29$ & $3.10$ \\ 
   \hline
   Milky Way & $12.04\pm0.12$ & $10.7\pm0.1$ & $9.9\pm0.1$ & $10.40\pm0.18$ & $1.65\pm0.19$ & $10.19\pm0.04$ & $0.30\pm0.06$ & $10.54\pm0.16$ & $2.5\pm0.4$  \\ 
   \hline
  \end{tabular} 
\end{table*}

\autoref{tab:sims} shows that the properties of the Milky Way are generally well reproduced by the suite of simulations. The halo mass, star-forming gas mass, non-star-forming gas mass, SFR, bulge mass bulge-to-total mass ratio, and disc scale radius all fall unambiguously within the range of values found in the \emosaics galaxies. The total stellar mass and the stellar disc mass of the Milky Way slightly exceed the largest value found in the simulations, but are still consistent with the covered range to within the uncertainties of the observations. \eagle is known to underestimate the stellar masses of Milky Way-mass dark matter haloes by up to a factor of two \citep[see figs.~4 and~8 of][]{schaye15}. Given that the stellar (disc) mass is the only marginally discrepant property of the simulated galaxies, we do not expect this to significantly influence the results presented in this work. None the less, it should be kept in mind when interpreting our findings. 

Across the sample, good analogues of the Milky Way (here defined as matching at least six out of nine properties to within a factor of two) are MW02, MW03, MW09, MW12, MW14, and MW23. However, the variety of properties listed in \autoref{tab:sims} and the lack of significant correlations between them also shows that a large enough sample of simulated galaxies should eventually host a perfect Milky Way analogue. Because the observed properties of the Milky Way are consistent with the range spanned by the 25 simulated galaxies, reproducing an exact copy of the Milky Way is therefore not required for the analysis performed in this work.

\section{The observed Galactic GC age-metallicity distribution} \label{sec:mwobs}

\begin{figure*}
\includegraphics[width=\hsize]{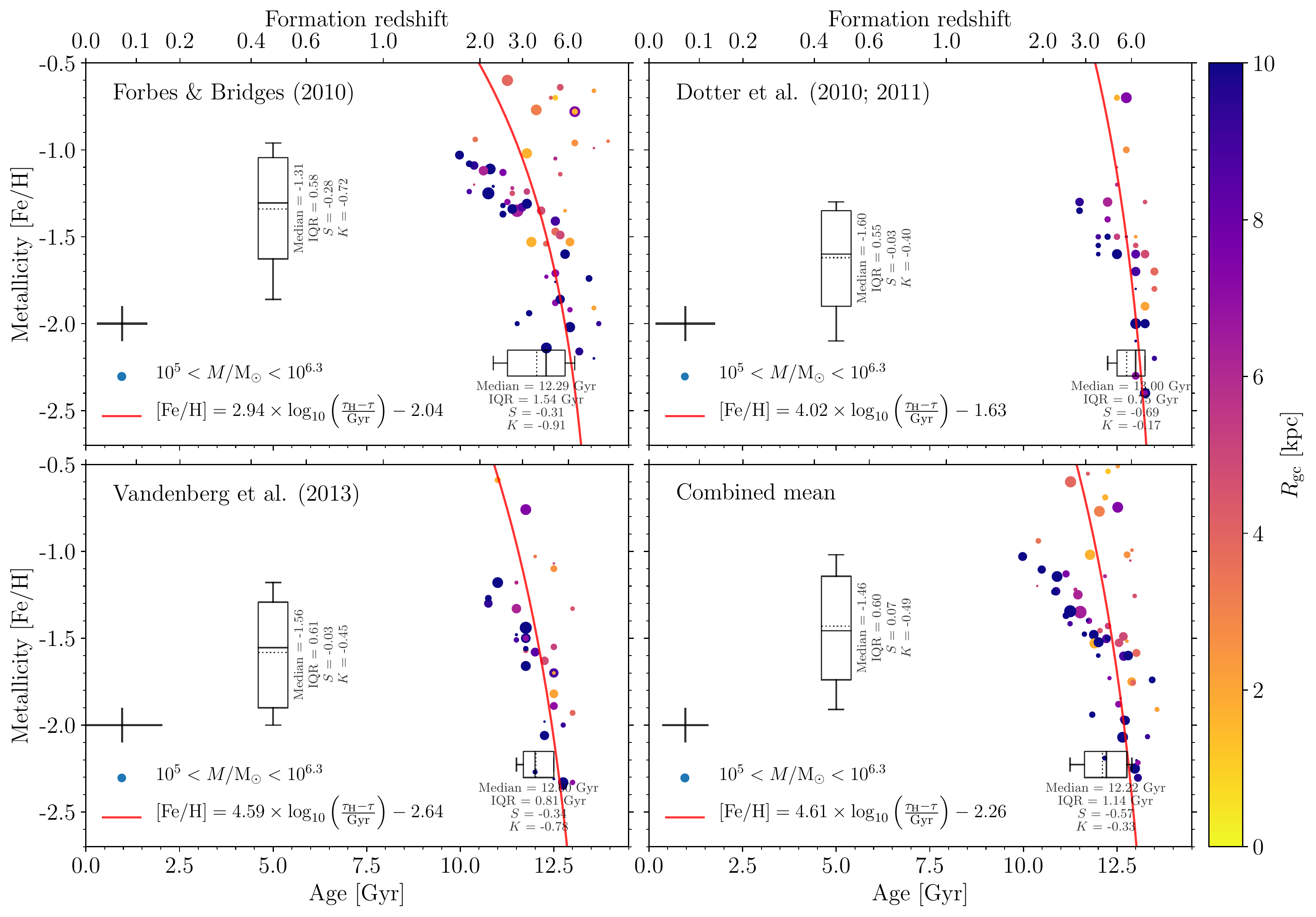}%
\caption{
\label{fig:agezmw}
Observed age-metallicity distributions of GCs in the Milky Way according to the compilations of \citet[top left panel]{forbes10}, \citet[top right panel]{dotter10,dotter11}, \citet[bottom left panel]{vandenberg13} and the combined mean of all three samples (bottom right panel). The symbol colour indicates the galactocentric radius according to the colour bar and the symbol size reflects the logarithm of the GC mass across the mass range indicated in each panel. Each age-metallicity distribution is characterised through two box plots and interquartile ranges of both age and metallicity, as well as by fitting the function shown in the bottom left corner of each panel (red line), with the best-fitting forms indicated (where $\thub=13.82~\gyr$ represents the age of the Universe). The various compilations are qualitatively similar, such that they can be used to constrain the formation and assembly history of the Milky Way.
}
\end{figure*}

\begin{table*}
  \caption{Quantities describing the GC age-metallicity distributions in the four (sub)samples of Galactic GCs. From left to right, the columns list: GC sample; median GC age $\widetilde{\tau}$ in Gyr; GC age interquartile range $\iqr(\tau)$ in Gyr; GC age skewness $S(\tau)$; GC age excess kurtosis $K(\tau)$; median GC metallicity $\widetilde{\feh}$; GC metallicity interquartile range $\iqr(\feh)$; GC metallicity skewness $S(\feh)$; GC metallicity excess kurtosis $K(\feh)$; combined interquartile range $\iqr^2\equiv\iqr(\tau)\times\iqr(\feh)$ in Gyr; interquartile range aspect ratio $\riqr\equiv\iqr(\feh)/\iqr(\tau)$ in Gyr$^{-1}$; best-fitting slope $\dfehdt$ of the function fitted in \autoref{fig:agezmw} indicating the rapidity of metal enrichment in the progenitor galaxies as traced by GCs; best-fitting intercept $\feh_0$ of the function fitted in \autoref{fig:agezmw} indicating the typical `initial' GC metallicity at $1~\gyr$ after the Big Bang; number of GCs (defined as clusters with present-day masses of $M>10^5~\msun$) in the metallicity range $-2.5<\feh<-0.5$ \citep[2010 edition]{harris96} that is considered throughout this work. The first four rows consider the observed GC samples from \autoref{fig:agezmw}, i.e.~from \citet[FB10]{forbes10}, \citet[D11]{dotter10,dotter11}, \citet[V13]{vandenberg13}, and the combined sample (see Appendix~\ref{sec:appobs}. To give an indication of the typical variation between these GC samples and the 25 simulated galaxies from \emosaics, the final rows list the minimum, median, maximum, and interquartile range of each column, both for the observed Galactic GC age-metallicity distribution and the modelled GC age-metallicity distributions in \emosaics.} 
\label{tab:gcmetricsobs}
  \begin{tabular}{l c c c c c c c c c c c c c}
   \hline
   Name & $\widetilde{\tau}$ & $\iqr(\tau)$ & $S(\tau)$ & $K(\tau)$ & $\widetilde{\feh}$ & $\iqr(\feh)$ & $S(\feh)$ & $K(\feh)$ & $\iqr^2$ & $\riqr$ & $\frac{{\rm d[Fe/H]}}{{\rm d}\log t}$ & $\feh_0$ & $\ngcp$ \\ 
   \hline
   \multicolumn{14}{c}{Observed Galactic GC age-metallicity distribution} \\
   \hline
   FB10 & $12.30$ & $1.54$ & $-0.31$ & $-0.91$ & $-1.31$ & $0.58$ & $-0.28$ & $-0.72$ & $0.90$ & $0.38$ & $2.94$ & $-2.04$ & $78$ \\ 
   D11 & $13.00$ & $0.75$ & $-0.69$ & $-0.17$ & $-1.60$ & $0.55$ & $-0.03$ & $-0.40$ & $0.41$ & $0.73$ & $4.02$ & $-1.63$ & $78$ \\ 
   V13 & $12.00$ & $0.81$ & $-0.34$ & $-0.78$ & $-1.56$ & $0.61$ & $-0.03$ & $-0.45$ & $0.49$ & $0.75$ & $4.59$ & $-2.64$ & $78$ \\ 
   Combined & $12.22$ & $1.14$ & $-0.57$ & $-0.33$ & $-1.46$ & $0.60$ & $0.07$ & $-0.49$ & $0.68$ & $0.52$ & $4.61$ & $-2.26$ & $78$ \\ 
   \hline
   Minimum & $12.00$ & $0.75$ & $-0.69$ & $-0.91$ & $-1.60$ & $0.55$ & $-0.28$ & $-0.72$ & $0.41$ & $0.38$ & $2.94$ & $-2.64$ & $78$ \\ 
   Median & $12.26$ & $0.97$ & $-0.46$ & $-0.55$ & $-1.51$ & $0.59$ & $-0.03$ & $-0.47$ & $0.59$ & $0.63$ & $4.31$ & $-2.15$ & $78$ \\ 
   Maximum & $13.00$ & $1.54$ & $-0.31$ & $-0.17$ & $-1.31$ & $0.61$ & $0.07$ & $-0.40$ & $0.90$ & $0.75$ & $4.61$ & $-1.63$ & $78$ \\ 
   IQR & $0.31$ & $0.44$ & $0.27$ & $0.52$ & $0.15$ & $0.03$ & $0.09$ & $0.12$ & $0.26$ & $0.25$ & $0.85$ & $0.42$ & $0$ \\ 
   \hline
   \multicolumn{14}{c}{Modelled GC age-metallicity distributions in \emosaics} \\
   \hline
   Minimum & $7.90$ & $0.44$ & $-3.15$ & $-1.16$ & $-1.59$ & $0.33$ & $-1.67$ & $-1.07$ & $0.22$ & $0.20$ & $2.33$ & $-3.90$ & $29$ \\ 
   Median & $10.73$ & $1.39$ & $-0.51$ & $0.63$ & $-0.91$ & $0.69$ & $-0.93$ & $-0.11$ & $1.06$ & $0.55$ & $3.50$ & $-2.69$ & $59$ \\ 
   Maximum & $12.45$ & $3.46$ & $0.71$ & $28.09$ & $-0.71$ & $0.88$ & $0.18$ & $2.18$ & $2.42$ & $1.11$ & $6.58$ & $-1.69$ & $187$ \\ 
   IQR & $1.01$ & $1.12$ & $1.79$ & $4.15$ & $0.12$ & $0.22$ & $0.44$ & $0.99$ & $0.73$ & $0.27$ & $1.32$ & $0.53$ & $72$ \\ 
   \hline
  \end{tabular} 
\end{table*}
The literature holds several compilations of the ages and metallicities of Galactic GCs \citep[e.g.][]{marinfranch09,forbes10,dotter10,dotter11,vandenberg13}. Here, we compile and combine three main compilations, i.e.~those by \citet{forbes10},\footnote{The compilation by \citet{forbes10} is based on original GC age measurements by \citet{salaris98}, \citet{bellazzini02}, \citet{catelan02}, \citet{deangeli05}, \citet{dotter08}, \citet{carraro07}, \citet{carraro09}, and \citet{marinfranch09}.} \citet{dotter10,dotter11}, and \citet{vandenberg13}. When combining these samples, we take the mean ages and metallicities whenever multiple measurements are available, resulting in a total of 96 GCs. A table listing these quantities and their uncertainties for the full sample of GCs is provided in Appendix~\ref{sec:appobs}. Before proceeding, we remove all GCs with masses $M<10^5~\msun$ and $\feh>-0.5$ to match the selection criteria used when analysing the \emosaics simulations in \citet{kruijssen18b}, leaving 61 GCs. This mass selection is made to minimise the effects of GC disruption, which is underestimated in \emosaics at lower GC masses in particular \citep[see][]{pfeffer18}, whereas the metallicity range is chosen to maximise the completeness of the available GC age measurements. For reference, the total number of Galactic GCs with $M\geq10^5~\msun$ (assuming $M_{V,\odot}=4.83$ and $M/L_V=2$) and $-2.5<\feh<-0.5$ is $\ngc=78$ \citep[2010 edition]{harris96}, showing that our sample contains 78~per~cent of the available GCs and is thus relatively complete. This completeness does not depend strongly on metallicity, as the sample includes $\sim70$~per~cent of the available GCs at $\feh\leq-1.5$ and $\sim90$~per~cent at $-1.5<\feh<-0.5$.

\autoref{fig:agezmw} shows the distributions of GCs from the four samples in age-metallicity space, including a set of quantitative metrics for describing the GC age-metallicity distribution derived in \citet[section~3.2; also see below]{kruijssen18b}. The age uncertainties are based on the original GC compilations. In addition, we adopt a fiducial metallicity uncertainty of $\sigma(\feh)=0.1~\dex$. Qualitatively, the GC age-metallicity distributions have a similar structure, with a steep branch at old ages, covering all metallicities, and a well-populated branch at intermediate metallicities that extends to young ages. The former branch includes mostly GCs at small galactocentric radii that are associated with the Galactic bulge, whereas the latter branch hosts mainly halo GCs, some of which are associated with tidal debris from satellite accretion. Indeed, in \citet{kruijssen18b} we show that the old branch mostly traces the in-situ growth of the main progenitor galaxy, whereas the young branch is constituted by GCs that formed ex-situ. Therefore, we refer to these branches as the `main branch' and the `satellite branch' respectively. \autoref{fig:agezmw} thus illustrates that the Milky Way acquired a significant part of its GC population through satellite accretion \citep[see e.g.][]{leaman13}.

There exist some quantitative differences between the four samples shown in \autoref{fig:agezmw}, most notably in their completeness and absolute age calibration, but there is no evidence for any strong systematic uncertainties. The bottom-right panel of \autoref{fig:agezmw} shows the combined mean catalogue and contains markedly less scatter around the satellite branch than each of the three individual samples. This suggests that the uncertainties are predominantly statistical rather than systematic, unless systematic errors affect all age measurements in the same way, and motivates using the combined mean sample when inferring the formation and assembly history of the Milky Way. We will therefore restrict our analysis to the `combined mean' GC sample throughout this paper.

Following section~3.2 of \citet{kruijssen18b}, we list 13 quantitative metrics describing the GC age-metallicity distributions in \autoref{tab:gcmetricsobs}. Several of these are also shown visually in \autoref{fig:agezmw}. The number of GCs $\ngcp$ refers to all Galactic GCs with present-day masses of $M>10^5~\msun$ (using $M_{V,\odot}=4.83$ and $M/L_V=2~\msun~{\rm L}_\odot^{-1}$) and metallicities $-2.5<\feh<-0.5$ listed in the \citet[2010 edition]{harris96} catalogue, irrespective of whether they are included in any of the listed literature samples. As listed, most of these metrics do not account for the uncertainties on the GC ages and metallicities. The only two exceptions are the age-metallicity slope $\dfehdt$ and intercept $\feh_0$, which have been obtained using an orthogonal distance regression \citep{boggs90} of the data with error bars. For the relevant subset of quantities, we obtain uncertainties in Section~\ref{sec:mwinf} below. The numbers listed in \autoref{tab:gcmetricsobs} confirm the visual impression from \autoref{fig:agezmw} that the GC age-metallicity distributions are qualitatively similar across all samples.

We can compare the metrics listed in \autoref{tab:gcmetricsobs} describing the GC age-metallicity distribution of the Milky Way to those of the 25 simulated galaxies in the \emosaics suite listed in table~2 of \citet{kruijssen18b}, of which the statistics are listed in the final four rows of \autoref{tab:gcmetricsobs}. The Milky Way GCs are old, with a somewhat narrow age distribution that has a normal skewness, but has underrepresented wings relative to its standard deviation (i.e.~a negative kurtosis), indicating an extended episode of GC formation over their full age range. The GC metallicities are lower than the median across the simulations,\footnote{This is possibly a result of the fact that GC disruption is underestimated in \emosaics due to the limited resolution (see \citealt{pfeffer18}). This manifests itself most strongly at low gas pressures, which are found predominantly in the low-redshift discs where metal-rich clusters are formed. As a result, \emosaics overpredicts the number of metal-rich GCs. However, this mostly occurs outside the metallicity range that is spanned by the observations and is considered here. The metallicity offset between the observations and simulations may also reflect a difference in calibration, or be caused by uncertainties on the nucleosynthetic yields in the simulations. Together, these introduce a systematic uncertainty of a factor of 2 \citep{schaye15}. Perhaps most importantly, the (absolute) median metallicity is not found to correlate with any galaxy-related quantities in \citet{kruijssen18b}, which means that it will not be used in the remainder of this work.} with considerably less absolute skewness than most of the simulated galaxies, but an otherwise ordinary metallicity distribution around the median. As a result, the Milky Way GC population is also relatively `average' in terms of the combined interquartile range $\iqr^2$ and the $\iqr$ aspect ratio $\riqr$. The best fits to the GC age-metallicity distribution are a bit steeper than average and attain a higher metallicity after $1~\gyr$ than most of the simulations do, indicating a relatively early collapse of the Galactic dark matter halo and rapid metal enrichment. Finally, the number of GCs is also slightly higher than average, with a median $\ngcp=\ngc/\fcorr=59$ for the simulations (where $\fcorr=1.75$ is a correction factor to account for the fact that GC disruption is underestimated in \emosaics, see the discussion in \citealt{kruijssen18b}).\footnote{In part, the somewhat lower number of GCs in the \emosaics galaxies relative to the Milky Way may be caused by the fact that \eagle underestimates the stellar masses of Milky Way-mass haloes by up to a factor of two (see the discussion in Section~\ref{sec:model}). Depending on when this additional stellar mass would have formed, it could plausibly have contributed a number of GCs. Accounting for this missing mass in the simulations could thus increase the number of GCs, at most by a factor of two.} Despite these small differences relative to the median properties of the simulated GC populations, we find that the ranges spanned by the simulations contain the Milky Way values without exception. This shows that the framework developed in \citet{kruijssen18b} for inferring the formation and assembly histories of galaxies from their GC populations can be applied to the Milky Way.

\section{Reconstructing the formation and assembly of the Galaxy} \label{sec:recon}

\subsection{Statistical description of the inferred formation and assembly history of the Milky Way} \label{sec:mwinf}

In \citet{kruijssen18b}, we systematically explore the correlations between the 13 quantities describing the GC age-metallicity distribution and a set of 30 quantities describing the formation and assembly of the host galaxy, for the entire sample of 25 Milky Way-mass galaxies in \emosaics. Out of the 390 correlations evaluated, 20 are found to carry a high statistical significance.\footnote{Additional correlations exist, but here we restrict our analysis to the strongest ones, because these are unambiguously physical in nature rather than a statistical coincidence. We emphasise that the relations adopted here are not the result of underlying dependences on the halo mass ($\mvir$) or other quantities against which possible correlations of the GC age-metallicity distribution are tested in \citet{kruijssen18b}. Specifically for $\mvir$, we find no statistically significant correlations due to the narrow range in halo mass covered by the \emosaics simulations. See \citet{kruijssen18b} for further details.} The best-fitting linear and power-law relations are listed in table~7 of \citet{kruijssen18b} and can be used to derive the galaxy-related quantities from the observed properties of the Galactic GC population. The 20 significant correlations rely on five GC-related quantities, i.e.~the median GC age $\widetilde{\tau}$, the GC age interquartile range $\iqr(\tau)$, the interquartile range aspect ratio $\riqr\equiv\iqr(\feh)/\iqr(\tau)$, the best-fitting slope of the GC age-metallicity distribution $\dfehdt$, and the number of GCs $\ngcp$ (as before, these are restricted to masses $M>10^5~\msun$ and metallicities $-2.5<\feh<-0.5$). Together, these metrics trace a total of 12 galaxy-related quantities. This is less than the 20 correlations, because some quantities are inferred through more than one GC-related metric.

\begin{figure*}
\center
\includegraphics[width=\hsize]{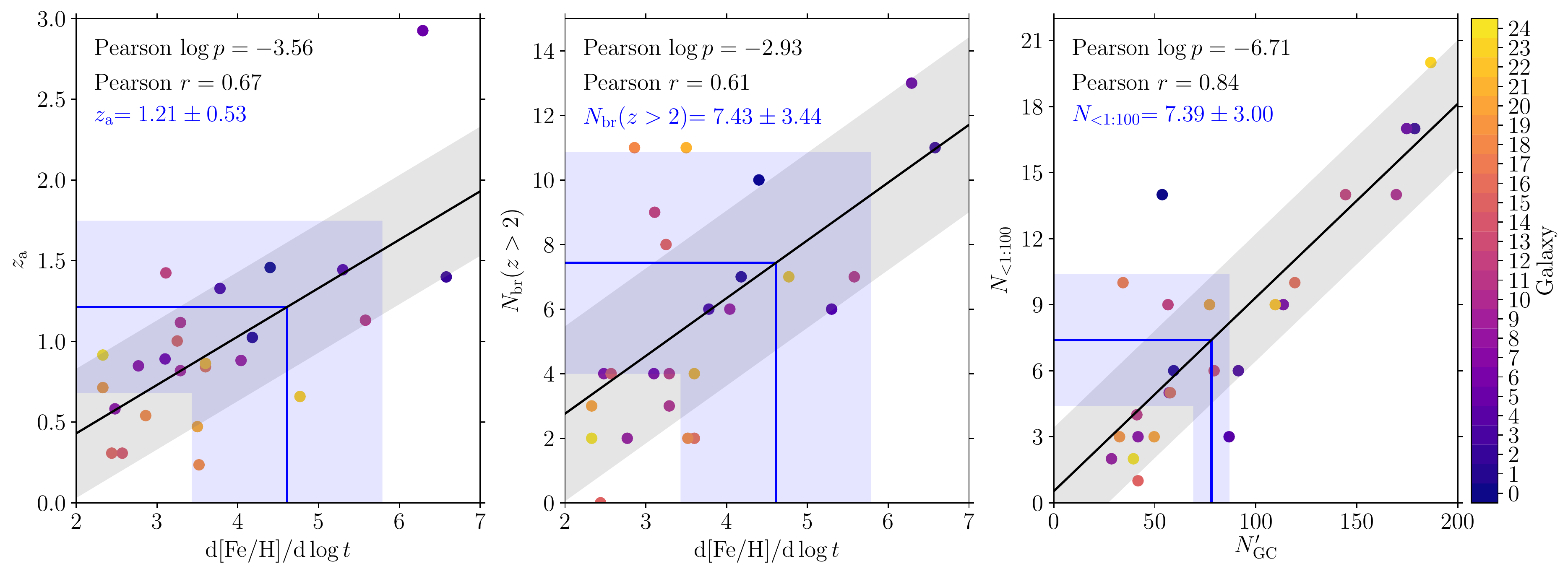}%
\caption{
\label{fig:correlationsmw}
Examples of three statistically significant correlations between galaxy formation and assembly-related quantities from \citet[section~4.2]{kruijssen18b} ($y$-axes) and the quantities describing the GC age-metallicity distribution from \citet[section~3.2]{kruijssen18b} ($x$-axes). Left panel: galaxy assembly redshift $\za$, which indicates when the main progenitor first attains 50~per~cent of the galaxy's stellar mass at $z=0$, as a function of the best-fitting slope of the GC age-metallicity distribution $\dfehdt$. Middle panel: number of high-redshift ($z>2$) mergers $\nbrz$ as a function of $\dfehdt$. Right panel: number of tiny mergers $N_{<1:100}$ (note that galaxies with stellar masses $M<4.5\times10^6~\msun$ are not counted as discrete mergers, but as smooth accretion), as a function of the (corrected) number of GCs $\ngcp$ (see the text) in the mass and metallicity range $M>10^5~\msun$ and $-2.5<\feh<-0.5$. Coloured symbols denote the 25 simulations as indicated by the colour bar on the right. In each panel, the solid line indicates the best-fitting linear regression, with the grey-shaded band marking the $1\sigma$ scatter of the data around the fit, and the Pearson $p$-values and correlation coefficients in the top-left corner indicating its statistical significance. The blue bands show how the quantities are inferred from the age-metallicity distribution of Galactic GCs in the bottom-right panel of \autoref{fig:agezmw}. Together, these place quantitative constraints on the formation and assembly history of the Milky Way.
}
\end{figure*}
The first set of six galaxy-related quantities describes the instantaneous properties and the mass growth history of the galaxy, i.e.~the maximum circular velocity $\vmax$, the concentration parameter of the dark matter halo profile $\cnfw$, the lookback times at which the main progenitor attains $\{25,50\}$~per~cent of its maximum halo mass $\{\ttf,\tfz\}$, the redshift at which the {\it main progenitor} reaches 50~per~cent of its final stellar mass $\za$, and the lookback time at which 50~per~cent of all stars currently in the galaxy have formed {\it across all progenitors} $\tf$.\footnote{These lookback times are chosen as `distance markers' along the Milky Way's continuous formation and assembly history, and deliberately deviates from the classical narrative of `two-phase' galaxy formation, which oversimplifies the complexity of galaxy assembly through a sharp separation between in-situ and ex-situ growth. Especially at the peak formation redshifts of GCs ($z=2$--$6$, \citealt{forbes15,reinacampos18b}), this distinction is highly ambiguous, because the intense merging activity at these redshifts can be an important driver of in-situ star formation by bringing gas into the main progenitor.} The second set describes the merger history and progenitor population of the galaxy, i.e.~the number of high-redshift ($z>2$) mergers experienced by the main progenitor $\nbrz$, its total number of mergers $\nbr$, the fraction of $z>2$ mergers $\rz$, the total number of progenitors $\nleaf$, the number of `tiny' mergers (with stellar mass ratio $<$1:100) $N_{<1:100}$, and the number of minor mergers (with stellar mass ratio 1:4--1:100) $N_{1:4-1:100}$. In this framework, `major' mergers are considered those with stellar mass ratios $>$1:4. Note that these quantities only include progenitors with stellar masses $M_*>4.5\times10^6~\msun$, to restrict the statistics of the progenitors to haloes hosting at least 20 stellar particles.

In order to determine the uncertainties of the obtained galaxy-related quantities, we must estimate the uncertainties on the five GC-related quantities that they are derived from. For $\widetilde{\tau}$, $\iqr(\tau)$, and $\riqr$, we obtain these by Monte Carlo sampling $10^5$ different realisations of the `combined mean' Galactic GC age-metallicity distribution (bottom-right panel of \autoref{fig:agezmw}), using the age and metallicity uncertainties of each individual GC and measuring the quantities for each realisation. The resulting distributions are close to Gaussian and we therefore adopt the standard deviations across the $10^5$ realisations as the uncertainties. For the age interquartile range $\iqr(\tau)$, which measures the age dispersion of the GC sample, we subtract the mean age error of the GCs in quadrature to avoid artificially boosting the underlying interquartile range. The resulting median metrics and their uncertainties are $\{\widetilde{\tau}, \iqr(\tau), \riqr\}=\{12.21\pm0.12,1.21\pm0.20,0.48\pm0.10\}$. These values are very similar to the numbers listed for the combined mean sample in \autoref{tab:gcmetricsobs}, which were derived without accounting for the uncertainties on the data. For the slope of the GC age-metallicity distribution $\dfehdt$, the uncertainties are obtained self-consistently from the weighted orthogonal distance regression to the data with their age-metallicity uncertainties, resulting in $\dfehdt=4.61\pm1.18$ (or $\log{\dfehdt}=0.664\pm0.112$). Finally, for the number of GCs, we assume Poisson errors, i.e.~$\ngcp=78\pm9$ (or $\log{\ngcp}=1.892\pm0.049$).

We now combine the above metrics describing the GC age-metallicity distributions with the best-fitting relations from table~7 of \citet{kruijssen18b} connecting them to several galaxy formation-related quantities to constrain the formation and assembly history of the Milky Way. \autoref{fig:correlationsmw} shows three examples, highlighting the relations between the slope of the GC age-metallicity distribution ($\dfehdt$) and the assembly redshift ($\za$, left) and number of high-redshift mergers ($\nbrz$, middle), as well as between the number of GCs ($\ngcp$) and the number of tiny mergers ($N_{<1:100}$, right). These relations show that the uncertainties are sufficiently small to provide meaningful constraints on the formation history of the Milky Way. \autoref{fig:correlationsmw} confirms the picture sketched in Section~\ref{sec:mwobs}, in which the Milky Way assembled early -- with $\za=1.2\pm0.5$, its fiducial assembly redshift (at which 50~per~cent of the $z=0$ stellar mass is first attained by the main progenitor) corresponds to the 76th percentile of the 25 simulations shown as coloured symbols.\footnote{The outlier at $\za\sim3$ is galaxy MW05, which quenches at $z\sim2$ and as a result has an assembly redshift biased to early cosmic times.} The number of high-redshift mergers $\nbrz$ is also larger than the median across the \emosaics suite (72th percentile), again indicative of an early formation and assembly phase. By contrast, the total number of tiny mergers implied by $\ngcp$ is $N_{<1:100}=7.4\pm3.0$, which is close to the median of our sample of simulated galaxies. It is thus not the number of accretion events experienced by the Milky Way that is elevated, but the lookback time at which these mergers took place that is larger than for most other galaxies.

\autoref{tab:fitmw} lists the complete set of metrics describing the formation and assembly history of the Milky Way that we obtain using the Galactic GC population. Reassuringly, the quantities for which we have multiple independent constraints all match to within the uncertainties. Physically, the combination of all quantities paints a consistent picture. Within our sample, the Milky Way has a typical mass and rotation curve, as indicated by the fact that the maximum circular velocity $\vmax=180\pm17~\kms$ implied by the number of GCs\footnote{This maximum circular velocity is inconsistent with the measured $\vmax$ of the Milky Way, but it matches the circular velocity at radii $R>40~\kpc$ in the Galactic halo \citep[e.g.][]{sofue15}. These radii are where $\vmax$ is typically found in the \emosaics galaxies. We interpret this as a sign of the known underestimation of the stellar masses of Milky Way-mass haloes in $\eagle$ (see the discussion in Section~\ref{sec:model}).} and the concentration parameter $\cnfw=8.0\pm1.0$ obtained from $\iqr(\tau)$, $\riqr$, and $\ngcp$ both correspond to the 48th percentiles of our simulations. 

The above picture changes when considering quantities that measure the growth rate of the Galaxy. The lookback times at which 25~and 50~per~cent of the Milky Way's final halo mass are predicted to have assembled, $\ttf=11.5\pm0.8~\gyr$ (i.e.~$z=3$) and $\tfz=9.4\pm1.4~\gyr$ (i.e.~$z=1.5$), correspond to the 76th and 72th percentiles of the \emosaics suite, respectively. For comparison, the application of the extended Press-Schechter formalism to the Planck cosmology adopted in this work shows that haloes of mass $\mvir=10^{12}~\msun$ reach \{25, 50\}~per~cent of their present-day mass at redshift $z=\{2.3, 1.2\}$ \citep{correa15b}. As with the above comparison to the \emosaics galaxies, this suggests that the Milky Way assembled faster than most galaxies of its halo mass. We have already seen in \autoref{fig:correlationsmw} that the same applies to the assembly redshift, which indicates when 50~per~cent of the final {\it stellar} mass has assembled into the main progenitor, and \autoref{tab:fitmw} extends this to the formation lookback time $\tf=10.1\pm1.4~\gyr$, which indicates the time at which 50~per~cent of the stellar mass at $z=0$ has been attained across all progenitors and corresponds to the 94th percentile of the sample of simulated galaxies. Comparing this to the equivalent number based on the star formation history from \citet{snaith14}, which is $\tf=10.5\pm1.5~\gyr$, we see that our estimate is in excellent agreement with the more direct,  star formation history-based measurement.
\begin{table}
\caption{Results of applying the 20 statistically significant correlations between quantities characterising galaxy formation and assembly and those describing the GC age-metallicity distribution to the GC population of the Milky Way. For each correlation, the table lists the predicted galaxy-related value based on the properties of the Galactic GC population, including the $1\sigma$ uncertainties. For those quantities that are constrained by multiple relations, the final column shows the weighted mean. See the text for details.}
\label{tab:fitmw}
\begin{tabular} {@{}ccccl@{}}
  \hline
 Quantity & [units] & Obtained from & Value & Combined \\ 
  \hline
  $\vmax$ & $[\kms]$ & $\log{\ngcp}$ &           $180\pm17$   &            \\
  $\cnfw$ & [--] & $\iqr(\tau)$ &                 $8.5\pm1.6$         & \hspace{-10pt}\rdelim\}{3}{10pt} \multirow{3}{*}{$8.0\pm1.0$}       \\
  $\cnfw$ & [--] & $\riqr$ &                        $7.6\pm1.7$         &         \\
  $\cnfw$ & [--] & $\log{\ngcp}$ &                     $7.9\pm1.8$        &          \\
  $\ttf$ & $[\gyr]$ & $\widetilde{\tau}$ &   $11.8\pm1.2$        & \hspace{-10pt}\rdelim\}{2}{10pt} \multirow{2}{*}{$11.5\pm0.8$}                             \\
  $\ttf$ & $[\gyr]$ & $\log{\dfehdt}$ &                $11.2\pm1.0$        &                          \\
  $\tfz$ & $[\gyr]$ & $\log{\dfehdt}$ &               $9.4\pm1.4$        &                           \\
  $\za$ & [--] & $\dfehdt$ &                       $1.2\pm0.5$         &          \\
  $\tf$ & $[\gyr]$ & $\widetilde{\tau}$ &    $10.1\pm1.4$        &                             \\
  $\nbrz$ & [--] & $\widetilde{\tau}$ &       $9.9\pm2.3$        & \hspace{-10pt}\rdelim\}{2}{10pt} \multirow{2}{*}{$9.2\pm1.9$}                         \\
  $\nbrz$ & [--] & $\dfehdt$  &                  $7.4\pm3.4$         &                       \\
  $\nbr$ & [--] & $\riqr$ &                          $14.1\pm6.1$       & \hspace{-10pt}\rdelim\}{3}{10pt} \multirow{3}{*}{$15.1\pm3.3$}        \\
  $\nbr$ & [--] & $\dfehdt$ &                     $18.9\pm7.4$       &             \\
  $\nbr$ & [--] & $\ngcp$ &                       $14.2\pm4.7$        &          \\
  $\rz$ & [--] & $\widetilde{\tau}$ &           $0.61\pm0.10$          &                     \\
  $\nleaf$ & [--] & $\log{\ngcp}$ &                     $24.1\pm10.2$      &              \\
  $N_{<1:100}$ & [--] & $\riqr$ &              $7.1\pm4.3$        & \hspace{-10pt}\rdelim\}{3}{10pt} \multirow{3}{*}{$7.9\pm2.2$}                  \\
  $N_{<1:100}$ & [--] & $\dfehdt$ &         $10.7\pm5.3$           &                       \\
  $N_{<1:100}$ & [--] & $\ngcp$ &            $7.4\pm3.0$         &                     \\
  $N_{1:100-1:4}$ & [--] & $\log{\ngcp}$ &       $4.4\pm2.3$         &                         \\
  \hline
\end{tabular}
\end{table}

The remainder of \autoref{tab:fitmw} lists a variety of quantities describing the merger history of the Milky Way. Firstly, we predict that the Milky Way experienced $\nbrz=9.2\pm1.9$ mergers before $z=2$. Even though the total number of mergers $\nbr=15.1\pm3.3$ experienced by the Milky Way is not unusually high (52nd percentile in \emosaics), the number of high-redshift mergers corresponds to the 80th percentile of the simulated galaxies. This directly implies a high fraction of mergers having taken place at $z>2$, coincident with the ages spanned by the Galactic GC population. Indeed, $\rz=0.61\pm0.10$ (88th percentile) suggests an early phase of rapid merging during which a large fraction of the Milky Way assembled. As shown in \citet{kruijssen18b}, the GC age-metallicity distribution does not strongly constrain the host galaxy's major merger activity (with stellar mass ratios $>$1:4), because the GC population is mainly shaped by minor mergers. For that reason, the final three quantities listed in \autoref{tab:fitmw} describe the minor merger history of the Milky Way. Before discussing these, we reiterate that any accreted satellites with stellar masses $M_*<4.5\times10^6~\msun$ are not counted as mergers, but as smooth accretion. The total number of progenitors $\nleaf=24.1\pm10.2$ (a fraction $1-\nbr/\nleaf\approx0.37$ of which merged prior to accretion onto the Milky Way) is not well-constrained due to the large uncertainties, but does correspond to a rather `normal' 48th percentile in \emosaics. Likewise, the number of tiny mergers ($N_{<1:100}=7.9\pm2.2$) and minor mergers ($N_{1:100-1:4}=4.4\pm2.3$) are common numbers across the simulation suite, corresponding to the 52th and 40th percentiles, respectively. Subtracting the number of tiny and minor mergers from the total number of mergers leaves space for a couple ($3\pm3$) of major mergers experienced by the Milky Way, which based on these numbers (e.g.~$\ttf=11.5\pm0.8~\gyr$) most likely took place during its early, rapid formation phase, i.e.~at $\tau>11~\gyr$ or $z>2.5$.

\subsection{The inferred merger tree of the Milky Way} \label{sec:mwtree}

\begin{figure*}
\includegraphics[width=0.91\hsize]{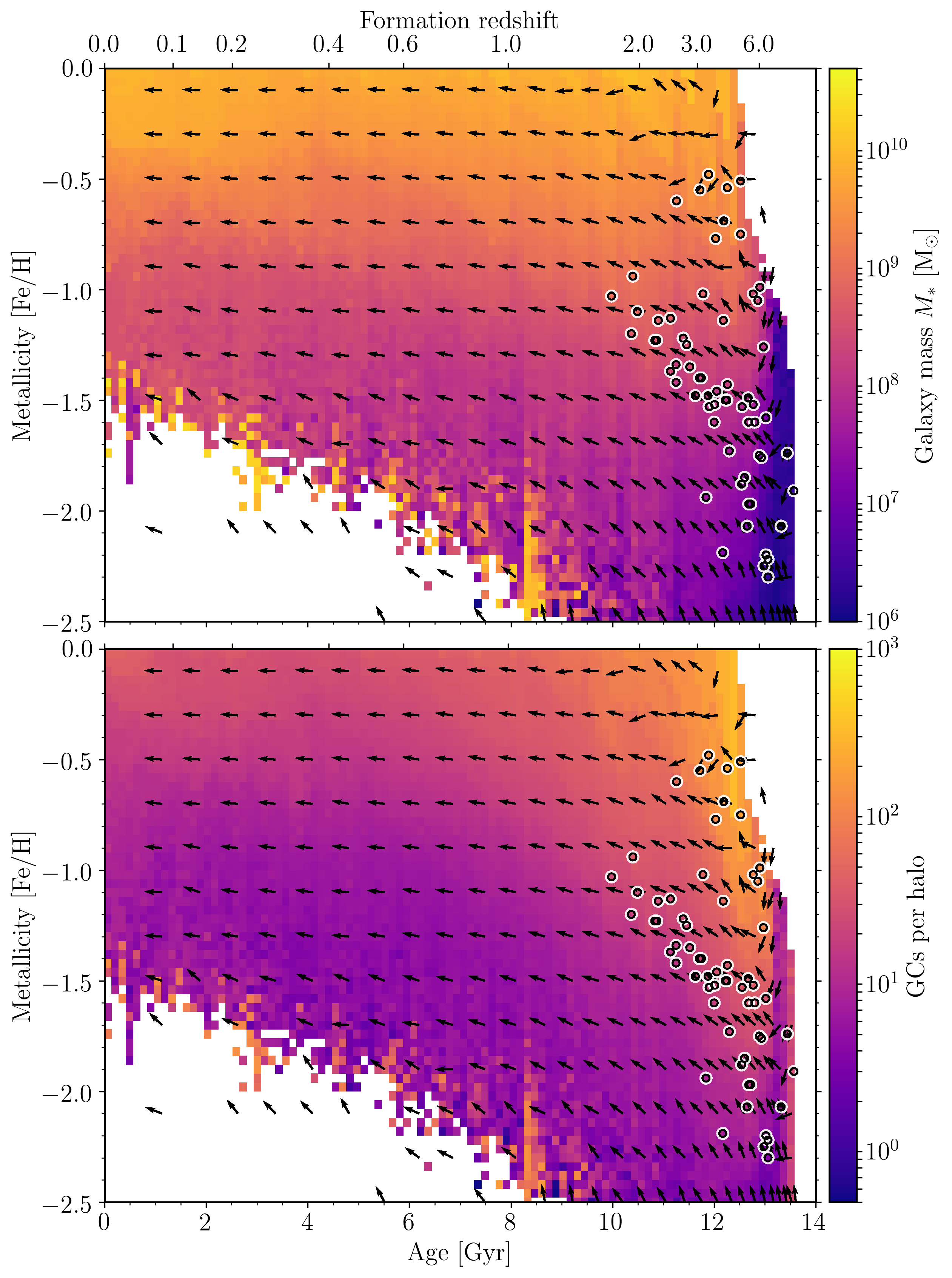}%
\caption{
\label{fig:agezmwmatch}
Age-metallicity evolution of star formation within the \eagle Recal-L025N0752 simulation for all galaxies with halo masses $\mvir\leq3\times10^{12}~\msun$. Top panels: at each age-metallicity coordinate, the colour indicates the median of the host galaxy stellar mass of all star particles formed since the previous snapshot. Bottom panels: the colour indicates the median number of surviving GCs expected to be hosted at $z=0$ by a galaxy forming stars at that age-metallicity coordinate (see the text). In both panels, vectors indicate the host's median metallicity evolution towards the next snapshot. The circles indicate the GC population of the Milky Way. In the text, we use the background colours in this figure to reconstruct the stellar mass growth history, metal enrichment history, and merger tree of the Milky Way.
}
\end{figure*}
In \citet{kruijssen18b}, we use the \emosaics simulations to develop a framework for reconstructing a galaxy's merger tree based on the age-metallicity distribution of its GC population. This approach links the ages and metallicities of the GCs to the evolution of galaxies in the phase space spanned by stellar mass, lookback time, and the metallicity of newly-formed stars, for all galaxies in the \eagle Recal-L025N0752 volume. Here, we apply this method to the GC population of the Milky Way. \autoref{fig:agezmwmatch} shows the GC age-metallicity distribution of the Milky Way, with the background colour indicating the expected host galaxy stellar mass of each GC (top panel) or the number of surviving GCs expected to be brought in by a galaxy forming stars at a given age-metallicity coordinate (bottom panel). The latter is estimated based on the galaxies' projected $z=0$ halo masses, the observed relation between the GC system mass and the dark matter halo mass at $z=0$ \citep[$M_{\rm GCs}/\mvir=3\times10^{-5}$,][]{durrell14,harris17}, and the mean GC mass at $M>10^5~\msun$, which is $\overline{M}=4.7\times10^5~\msun$ for a typical GC mass function \citep[e.g.][]{jordan07}. This results in $\ngcp=10^{-10.2}\times[\mvir(z=0)/\msun]$. For reference, the vector field indicates the typical evolution of galaxies in age-metallicity space. By connecting the Galactic GCs along such flow lines to the background colour, we can infer the properties of the Milky Way's progenitors and reconstruct part of its merger tree. The top panel clearly demonstrates that GC metallicity is the strongest indicator of the mass of its host galaxy at formation, with a secondary dependence on GC age.

The main branch of the Galactic GC population shown in \autoref{fig:agezmwmatch} is very steep, straddling the edge of the age-metallicity range occupied by the simulated galaxies in the \eagle Recal-L025N0752 volume. This suggests rapid stellar mass growth and metal enrichment, which is consistent with the findings from Section~\ref{sec:mwinf}. As discussed in \citet{kruijssen18b}, the stellar mass growth and metal enrichment histories of the main progenitor can be read off \autoref{fig:agezmwmatch} by following the upper envelope of the GC population's main branch and subtracting $\Delta\feh=0.3~\dex$.\footnote{This downward metallicity correction is recommended in \citet{kruijssen18b} to account for the fact that GCs have slightly higher metallicities than their host due to being born in the central regions of their host galaxy.} Given the age range of the GC population of the main branch, this is possible for redshifts $z=3$--$6$. The result is listed in \autoref{tab:prog}, together with the expected number of GCs with mass $M>10^5~\msun$ hosted by the Galaxy at $z=0$ as obtained from the bottom panel in \autoref{fig:agezmwmatch}.\footnote{The expected number of GCs has asymmetric uncertainties. At a given age-metallicity coordinate, the \eagle Recal-L025N0752 volume contains a non-negligible number of host galaxies that end up with halo masses at $z=0$ that are considerably lower than the median, which would therefore hold proportionally fewer GCs. These downwards uncertainties are typically a factor of 2 and the number of GCs in \autoref{fig:agezmwmatch} should be read with this caveat in mind. Therefore, we use the Poisson error for upward uncertainties and a factor of 2 as the downward uncertainties on $\ngcp$.} In addition to these numbers, the correlation analysis from Section~\ref{sec:mwinf} and \autoref{tab:fitmw} revealed that 50~per~cent of the Milky Way's stellar mass at $z=0$ (which is $M_*=5\pm1\times10^{10}~\msun$, see e.g.~\citealt{blandhawthorn16}) should have assembled into the main progenitor by $z=1.2\pm0.5$. We find that \eagle galaxies reaching $M_*=2.5\times10^{10}~\msun$ at that redshift have typical star	-forming gas metallicities of $\feh=0.1$. This combination of age, mass, and metallicity thus provides an additional data point in the formation and assembly history of the Milky Way, even if our sample does not contain any GCs with such young ages.
\begin{table}
\caption{Stellar mass growth and metal enrichment histories of the four identified (and likely most massive) Milky Way progenitors, identified using the age-metallicity distribution of Galactic GCs. We also list the number of GCs with masses $M>10^5~\msun$ expected to be hosted by each progenitor were it to survive as an undisrupted galaxy until $z=0$. The uncertainties on the inferred stellar masses and metallicities are $0.26~\dex$ and $0.25~\dex$, respectively \citep[section~5.3]{kruijssen18b}, unless stated otherwise.}
\label{tab:prog}
\begin{tabular} {@{}rccccl@{}}
  \hline
 Progenitor & $z$ & $\tau/\gyr$ & $\log{M_*/\msun}$ & $\feh$ & $\ngcp$ \\ 
  \hline
  \multirow{4}{*}{Main} \ldelim\{{4}{0pt} & 6 & $12.9$ & $8.2$ &                 $-1.3$         & \hspace{-10pt}\rdelim\}{4}{10pt} \multirow{4}{*}{$150_{-75}^{+12}$}       \\
   & 4 & $12.3$ & $9.1$ &                        $-0.8$         &         \\
   & 3 & $11.7$ & $9.5$ &                        $-0.5$         &         \\
   & $1.2\pm0.5$ & $8.7_{-2.2}^{+1.3}$ & $10.4_{-0.1}^{+0.1}$ &                        $0.1$         &         \\[1ex]\cline{2-6}\rule{0pt}{3ex} 
  \multirow{5}{*}{Sat~1 \& 2} \ldelim\{{5}{0pt} & 6 & $12.9$ & $6.9$ &                 $-1.8$         & \hspace{-10pt}\rdelim\}{5}{10pt} \multirow{5}{*}{$22_{-11}^{+5}$}       \\
   & 4 & $12.3$ & $8.0$ &                 $-1.6$         &  \\
   & 3 & $11.7$ & $8.2$ &                        $-1.4$         &         \\
   & 2.3 & $11.0$ & $8.6$ &                        $-1.2$         &         \\
   & 1.7 & $10.0$ & $9.1$ &                        $-1.0$         &         \\[1ex]\cline{2-6}\rule{0pt}{3ex} 
  \multirow{3}{*}{Sat~3} \ldelim\{{3}{0pt} & 6 & $12.9$ & $6.3$ &                 $-2.4$         & \hspace{-10pt}\rdelim\}{3}{10pt} \multirow{3}{*}{$8_{-4}^{+3}$}       \\
   & 4 & $12.3$ & $7.0$ &                        $-2.2$         &         \\
   & 3 & $11.7$ & $7.6$ &                        $-1.9$         &         \\
  \hline
\end{tabular}
\end{table}

Similar constraints can be placed on the stellar mass evolution and metal enrichment of the accreted satellites, using the GC satellite branch in \autoref{fig:agezmwmatch}. In \citet{kruijssen18b} we claimed that, as a general guideline based on our 25 cosmological zoom-in simulations in \emosaics, the properties of the accreted satellites can be inferred for $z=1$--$2.5$ (or $\tau=7.9$--$11.2~\gyr$). However, owing to the rapid growth of the Milky Way's main progenitor, the GC satellite branch separates from the main branch already at a considerably higher redshift ($z\sim5$), enabling the satellite population to be followed over an unusually broad redshift range of $z=1.7$--$6$ (or $\tau=10$--$12.9~\gyr$) thanks to the vector fields shown in \autoref{fig:agezmwmatch}. The Milky Way's satellite branch is quite wide at $z=3$--$4$ (or $\tau\sim12~\gyr$), spanning $\Delta\feh=0.9~\dex$ from $\feh=-2.2$ to $\feh=-1.3$. This implies the existence of multiple accreted satellites of various masses. About 35 GCs are associated with a single satellite branch that extends to $z=1.7$ and relatively high metallicities of $\feh=-1.0$.\footnote{This number includes half of the GCs at the oldest ages and lowest metallicities, where the satellite and main branches merge. In \citet{kruijssen18b}, we show that the GCs in that area of age-metallicity space are typically equally split between the main progenitor and its satellites.} Using the bottom panel of \autoref{fig:agezmwmatch}, we find that galaxies forming stars at these ages and metallicities are expected to host $22_{-11}^{+5}$ GCs by $z=0$. Because star formation in these progenitors ceased prior to $z=0$ due to their accretion onto the Milky Way, this number of GCs is technically an upper limit. However, for the progenitors in which the youngest GCs on the satellite branch formed, we do not expect the true number to be much lower -- the presence of these young GCs shows that the time of accretion must have been at $z<1.7$, well after the peak of GC formation activity \citep{reinacampos18b}.

In view of the number of GCs expected per satellite ($22_{-11}^{+5}$) and the number actually hosted by the satellite branch ($\sim35$), it is most likely that the bulk of the satellite branch was populated by two accreted galaxies of similar mass, which each brought in $\sim15$ GCs. The presence of two such satellites may partially explain the large metallicity spread of the satellite branch. However, we cannot identify a metallicity bimodality in this branch, implying that it is impossible to distinguish between the satellites. We therefore assign the same properties to `Satellite~1' and `Satellite~2' across the redshift range $z=1.7$--$6$. Finally, at the low end of the mass and metallicity range ($\{\tau/\gyr,\feh\}=\{12,-2\}$), the GCs on the satellite branch extend to considerably lower metallicities than the median metallicity. We associate these with the lowest-mass satellite that we can identify. This `Satellite~3' can be followed for a rather narrow redshift range using the vectors in \autoref{fig:agezmwmatch}, from $z=3$--$6$. It may have hosted up to five massive ($M>10^5~\msun$) GCs, depending on how many GCs at the oldest ages and lowest metallicities are associated with this satellite. For comparison, the expected number based on \autoref{fig:agezmwmatch} is $8_{-4}^{+3}$, which is consistent with the number of GCs possibly associated with Satellite~3. For all three identified satellites, the stellar mass growth and metal enrichment histories inferred from \autoref{fig:agezmwmatch} are listed in \autoref{tab:prog}.

\begin{figure*}
\includegraphics[width=0.79\hsize]{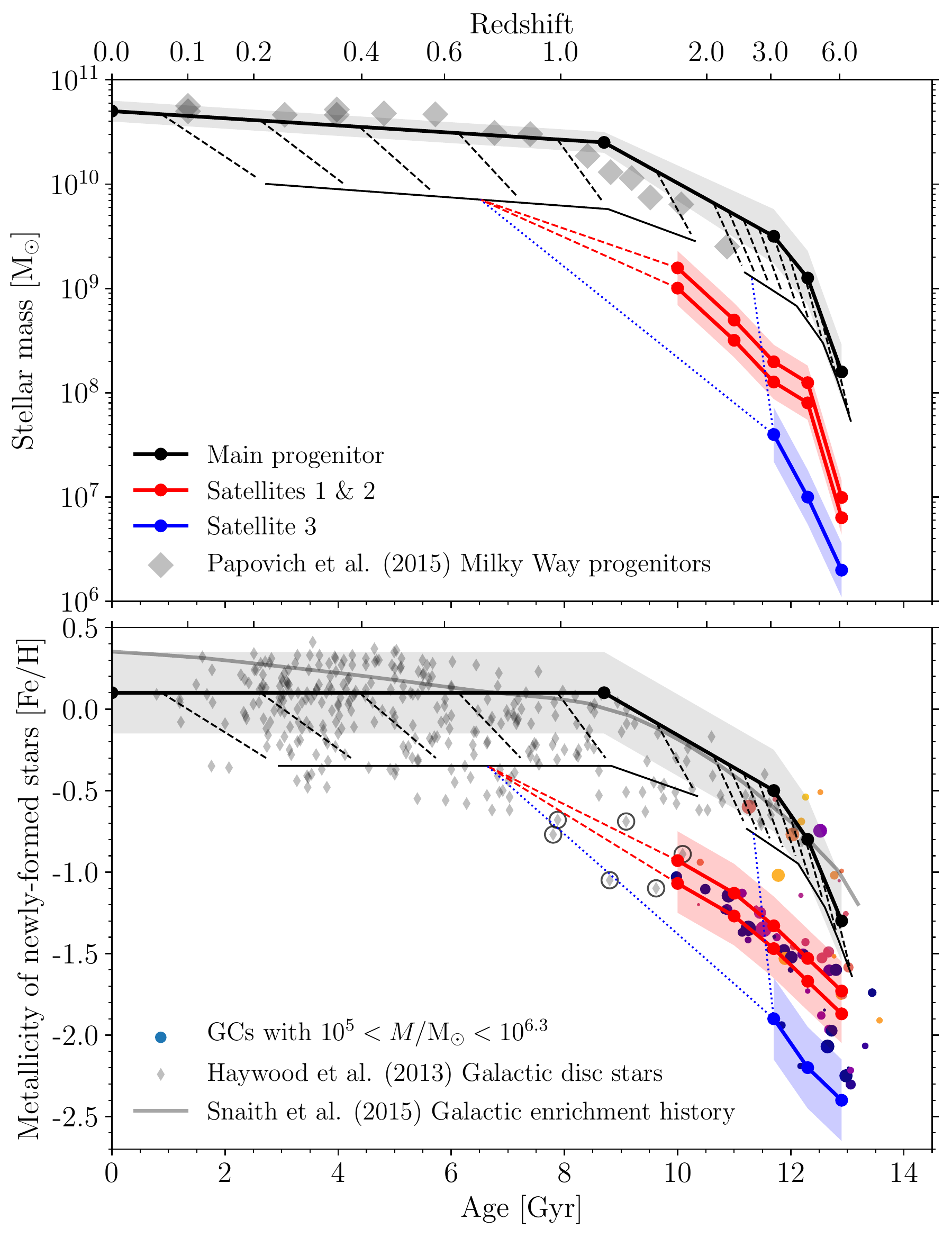}%
\caption{
\label{fig:mwevo}
Stellar mass growth (top panel) and metal enrichment history (bottom panel) of the Milky Way and three of its main satellites, inferred using the observed distribution of the Galactic GC population in age-metallicity space (see \autoref{tab:fitmw} and \autoref{tab:prog}). The thick solid lines indicate the progenitors as indicated by the legend in the top panel -- to distinguish Satellites~1 and~2, we have added a small vertical offset to the red lines. The shaded areas indicate the uncertainties around the inferred masses and metallicities. Thin lines sketch the topology of the Milky Way's merger tree, projected into age-mass (top) and age-metallicity (bottom) space. These lines have no meaning in terms of mass growth or enrichment and use the information from \autoref{tab:fitmw} that the Milky Way experienced $\nbrz\sim9$ mergers at $z>2$ and $\nbr\sim15$ mergers in total (implying $\sim6$ mergers at $z<2$). The mergers into the main branch are represented by dashed lines, which are spaced equally over the time intervals $z<2$ and $z>2$. Topologically, these mergers are a robust result of our analysis, even if their exact timing within each redshift interval is unknown. The three satellites are connected to the grouped accretion events of both time intervals (thin solid lines) by dashed and dotted lines, with the latter indicating that it is unknown whether Satellite~3 accreted before or after $z=2$. The large grey diamonds in the top panel show the stellar mass growth histories of Milky Way progenitor galaxies from \citet{papovich15}, which on average undergo less rapid growth than the Milky Way. For reference, the coloured symbols in the bottom panel match the Galactic GC population in the bottom-right panel of \autoref{fig:agezmw}, using the same colour coding of the galactocentric radius. In addition, the small grey diamonds in the bottom panel show the age-metallicity measurements for Galactic disc field stars from \citet{haywood13}, of which six stars that may have formed in Satellites 1--3 are marked with black circles (see Section~\ref{sec:known}). The grey line represents the fiducial metal enrichment history of \citet{snaith15}, which is based on the observed star formation history of the Milky Way. This figure demonstrates that the inferred formation and assembly history of the Milky Way is consistent with several independent sets of observational data, revealing that the Galaxy assembled through a combination of rapid in-situ growth and the accretion of several low-mass ($M_*\sim10^8$--$10^9~\msun$) satellite galaxies.
}
\end{figure*}
\autoref{fig:mwevo} visualises the inferred stellar mass growth histories and metal enrichment histories of the identified progenitors, as well as the merger tree projected in age-mass and age-metallicity space. The rapid assembly of the main progenitor is illustrated by the fact that 10~per~cent of its final mass is attained at $z=2.5$, and 50~per~cent at $z=1.2$. For comparison, the median Milky Way-mass galaxy in the volume-limited sample of 25 \emosaics simulations reaches 50~per~cent of its final mass at $z=0.9$ \citep{kruijssen18b}. Abundance matching studies predict that haloes of mass $\mvir=10^{12}~\msun$ reach \{10, 50\}~per~cent of their present-day stellar mass at redshift $z=\{1.9, 0.9\}$ \citep{behroozi13,moster13}. This rapid mass growth of the Milky Way is also found by comparison to the observed stellar mass growth histories obtained for Milky Way progenitors between redshifts $z=0.1$--$2.5$ \citep[grey diamonds in the top panel of \autoref{fig:mwevo}]{papovich15}. At redshifts $1<z<2.5$, the Milky Way's stellar mass growth proceeds $\Delta\tau=0.6$--$0.9~\gyr$ ahead of observed galaxies that are expected to reach the same stellar mass at $z=0$. This again implies that the Milky Way assembled its stellar mass considerably faster than most galaxies of the same halo mass.

\autoref{fig:mwevo} also shows the basic geometry of the merger tree connecting the four progenitors. Following \autoref{tab:fitmw}, the Milky Way experienced $\sim9$ mergers at $z>2$, leaving $\sim6$ mergers at $z<2$ (given its total number of $\sim15$ mergers). Satellites~1 and~2 contain GCs with formation redshifts $z<2$, hence they must have been part of the $\sim6$ mergers at $z<2$. However, their exact time of accretion is unconstrained, as this may have occurred at any time since the formation of the youngest GC on the satellite branch. The accretion time of Satellite~3 is even more poorly constrained based on the age-metallicity distribution only, because it hosts GCs with a narrow range of formation redshifts ($z=3$--$4$). Therefore, it may have merged into the main branch at any time $z<3$. We refine this statement in Section~\ref{sec:known}.

At fixed lookback time or redshift, the identified satellites have stellar masses that are 1--2 orders of magnitude below the main branch, with their most recent masses spanning $M_*=10^8$--$10^9~\msun$. Together with the steep main branch in age-metallicity space, this suggests that the Milky Way assembled through a combination of vigorous in-situ growth and satellite accretion, without any major mergers since $z\sim4$. At $z<4$, any galaxy with a stellar mass comparable to the Milky Way's main progenitor (i.e.~mass ratio $>$1:4) would have contributed more GCs and at higher metallicities than can be accommodated by the main branch and the satellite branch of the GC age-metallicity distribution. Due its unstructured nature at higher redshifts, the GC age-metallicity distribution does not enable the identification of major mergers at $z>4$.

Even though the accretion events of the three satellites onto the main progenitor were only minor mergers, their GC populations do stand out most clearly in \autoref{fig:agezmw} and \autoref{fig:agezmwmatch}. This suggests that the satellites were the most massive to have been accreted by the Milky Way. Lower-mass satellites must have been accreted too, given the expected $\nbr=15.1\pm3.3$ listed in \autoref{tab:fitmw}, but these likely had such low masses that some of them contributed only a couple of GCs per satellite ($\sim4$ satellites, see Section~\ref{sec:comp}), whereas most of them hosted no GCs at all ($\sim8$ satellites).

The bottom panel of \autoref{fig:mwevo} demonstrates that the inferred properties of the Milky Way's progenitors are consistent with several independent observables. Firstly, they trace the age-metallicity distribution of Galactic GCs. This is by construction, but the figure also highlights that the identified progenitors together cover most of the age-metallicity space spanned by the GCs. Secondly, the inferred metal enrichment history of the main progenitor closely matches the age-metallicity distribution of Milky Way disc stars from \citet{haywood13}, showing that the GCs and field stars in a galaxy follow similar age-metallicity distributions, consistent with our prediction from the \emosaics simulations \citep{kruijssen18b}.\footnote{The \citet{haywood13} field stars reside in the solar neighbourhood, within a distance of $\sim100~\pc$. However, this does not mean these stars formed there. The dynamical redistribution of stars by radial migration \citep[e.g.][]{sellwood02} or satellite accretion implies that they are expected to have originated from a variety of environments. This has no implications for the comparison in~\autoref{fig:mwevo}, as we merely include the stellar sample with the goal of comparing the enrichment histories of the Milky Way's progenitors inferred from its GCs to a typical field star sample that currently resides in the Galactic disc.} Thirdly, the expected metal enrichment history from the observed star formation history of the Milky Way from \citet{snaith15} traces the enrichment history of the main progenitor that we infer from the Galactic GC population nearly perfectly (i.e.~within $\Delta\feh=0.1~\dex$) over the redshift interval $z=0.4$--$4.5$. These enrichment histories have been obtained through completely independent methods, implying that their agreement provides an important validation of our results. Finally, even if it is not shown in \autoref{fig:mwevo}, the field star age-metallicity relations of the Sagittarius dwarf and the GCs formerly associated with Canis Major (both of which have recently been accreted by the Milky Way and possibly correspond to Satellites~2 and~3, see Section~\ref{sec:known}) are consistent with the enrichment histories derived for Satellites~2 and~3 \citep[see e.g.][]{layden00,forbes04,siegel07}. Together, these independent lines of evidence strongly suggest that the properties of the Milky Way's progenitors inferred from their GC population are robust and carry predictive power.

\section{Discussion} \label{sec:disc}

\subsection{Connection of results to known satellites and GCs} \label{sec:known}

\begin{figure}
\includegraphics[width=\hsize]{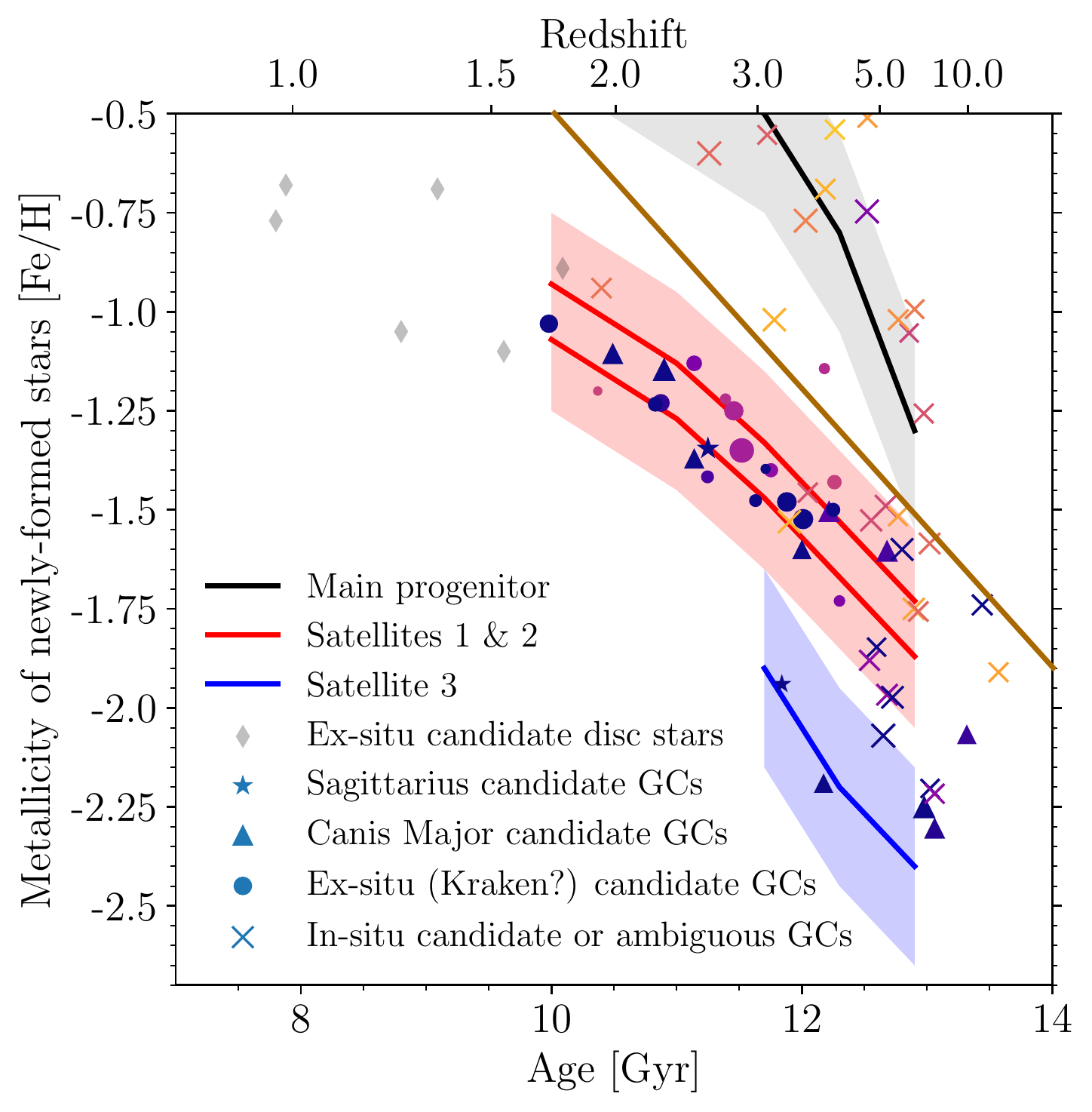}%
\caption{
\label{fig:mwevo_zoom}
Zoom-in of the age-metallicity distribution of Galactic GCs, highlighting their possible origin. The coloured symbols match the Galactic GC population in the bottom-right panel of \autoref{fig:agezmw}, using the same colour coding of the galactocentric radius. However, this time the symbol shape refers to the GCs' origin, as indicated by the legend. In addition, the small grey diamonds indicate the six Galactic disc field stars that may have formed in Satellites 1--3. As in \autoref{fig:mwevo}, the solid lines indicate the metal enrichment histories of the progenitors, with shaded areas indicating the uncertainties around the inferred metallicities. This figure illustrates that we expect $\sim2$ massive ($M>10^5~\msun$) GCs to have been brought in by Sagittarius, $\sim10$ by the `Canis Major' event, and 10--20 by a new, massive satellite accreted $6$--$9~\gyr$ ago, which we name {\it Kraken}.
}
\end{figure}

We identify three massive ($M_*=10^8$--$10^9~\msun$) satellite progenitors that have been accreted by the Milky Way since $z\sim3$, based on the properties of the satellite branch in the age-metallicity distribution of Galactic GCs. Several of these GCs are spatially and kinematically associated with known satellites of the Galaxy.\footnote{Recall that we restrict our analysis to GCs with masses $M>10^5~\msun$.} The most prominent of these satellites are the Sagittarius dwarf \citep[NGC5634 and NGC6715, see][]{bellazzini02,bellazzini03,law10,forbes10} and the GCs formerly associated with the Canis Major `dwarf' \citep[NGC~1851, NGC~1904, NGC~2808, NGC~4590, NGC~5286, NGC~6205, NGC~6341, NGC~6779, NGC~7078, IC~4499,\footnote{\label{ft:cma}The inclusion of several of these GCs hinges on their association with the Monoceros stream, which \citet{martin04} proposed is the tidal debris of a dwarf galaxy named Canis Major. However, the current consensus is that the Canis Major dwarf never existed (see \citealt{penarrubia05}, \citealt{deboer17}, \citealt{deason18}, and \citealt{carballobello18} for discussions). We do note that all of these GCs except NGC~4590, NGC~7078, and IC~4499 have highly similar apocentre distances, pericentre distances, eccentricities, and angular momenta \citep{helmi18,sohn18}. For historical reasons, we therefore maintain the nomenclature of referring to one of the identified progenitor satellites as Canis Major, but acknowledge that the associated GCs were brought in by a different progenitor.} see][]{martin04,penarrubia05,carballobello18}, which are currently undergoing tidal stripping and accretion onto the Galaxy \citep{ibata94,majewski03,martin04}. \autoref{fig:mwevo_zoom} shows a zoom-in of the GC age-metallicity distribution from the bottom panel of \autoref{fig:mwevo}, highlighting in which (accreted) satellites the GCs may have formed by using different symbol shapes. In this figure, GCs unassociated with any of the satellites (see below) are considered to be ambiguous in terms of having an in-situ or ex-situ origin. In addition, GCs with metallicities $\feh>2-0.25\tau/\gyr$ are considered to have formed in-situ, unless they have been spatially or kinematically associated with a satellite in the literature. The field star age-metallicity relation of Sagittarius connects directly to the satellite branch of the GC age-metallicity distribution, albeit to the lower end of its metallicity range \citep[e.g.][]{layden00,siegel07}, and the old open cluster population of the `Canis Major' GCs is a natural continuation of the satellite branch towards younger ages and higher metallicities \citep{forbes04}. It is therefore plausible that one of the identified satellites represents Sagittarius and another corresponds to the galaxy that contributed the `Canis Major' GCs.

Based on the number of associated GCs, we expect Sagittarius to have been the least massive of the above two known accreted satellites. In the age-metallicity distribution of Galactic GCs shown in \autoref{fig:mwevo_zoom}, NGC~5634 ($\{\tau/\gyr,\feh\}=\{11.84\pm0.51,-1.94\}$) is associated with Satellite~3. Due to its high metallicity, NGC~6715 ($\{\tau/\gyr,\feh\}=\{11.25\pm0.59,-1.34\}$) resides in the main satellite branch and thus coincides with Satellites~1 and~2 in age-metallicity space. However, this GC most likely represents the nuclear cluster of Sagittarius, implying an extended star formation history and a wide metallicity distribution, extending down to $\feh=-1.9$ \citep{carretta10}. This lower bound makes NGC~6715 a natural extension of the Satellite~3 age-metallicity track. Also in terms of its number of massive GCs ($\ngc\ga2$) and its estimated total number of GCs \citep[$\ngc=5$--$10$,][]{law10}, Sagittarius is likely to have represented the lowest-mass progenitor among the three that we identified, for which we estimate $\ngc=8_{-4}^{+3}$ had it evolved to $z=0$ without being disrupted. This match is corroborated by Sagittarius' estimated stellar mass prior to accretion onto the Milky Way, which is several $10^8~\msun$ \citep[e.g.][]{niedersteostholt10,niedersteostholt12}. Our predicted early ($z\geq3$) stellar mass growth and metal enrichment history is also consistent with these findings -- with a mass and metallicity of $M_*=4\times10^7~\msun$ and $\feh=-1.9$ at $z=3$, Satellite~3 is expected to reach a mass of $M_*\sim2\times10^8~\msun$ at $z=1$ \citep[e.g.][]{moster13,behroozi13}, with an expected metallicity of $\feh=-0.9$ \citep[e.g.][]{forbes10}. This projected metallicity is $\Delta\feh\sim0.3~\dex$ lower than the metallicity of the young ($\tau=3$--$8~\gyr$), low-mass ($M<10^5~\msun$) GC population of Sagittarius \citep{law10}, confirming the known result that Sagittarius has a current metallicity higher than expected for its relatively low mass \citep[e.g.][]{martinezdelgado05}. In view of these results, it is likely that Satellite~3 is the Sagittarius dwarf.

Progenitors more massive than Sagittarius are required for having populated the satellite branch of the Milky Way's GC age-metallicity distribution. As discussed above, we predict the existence of two such progenitors, dubbed Satellite~1 and~2. With its substantial population of 5--10 potential, massive GCs, the galaxy that contributed the `Canis Major' GCs is the obvious candidate for representing one of these satellites. \autoref{fig:mwevo_zoom} shows that all `Canis Major' GCs with formation redshifts $z\leq6$ (or $\tau>12.9~\gyr$) match the age-metallicity track of Satellite~1 and~2. Contrary to Sagittarius, its putative nuclear cluster NGC~2808 \citep{forbes10,carballobello18} does not have a metallicity spread \citep{carretta15} and as a result it traces the age-metallicity evolution of its host at $\{\tau/\gyr,\feh\}=\{10.90\pm0.57,-1.14\}$. The expected number of massive GCs ($\ngc=22_{-11}^{+5}$) provides a good match to the observed number, provided that we missed one or two associated GCs that should be added to the 10 `Canis Major' GCs. In terms of its stellar mass prior to accretion onto the Milky Way, \citet{penarrubia05} obtain $M_*=6\pm3\times10^8~\msun$, with the caveat that the mass is poorly constrained and could be as high as $M_*=2\times10^9~\msun$. Such a high mass would be consistent with the mass evolution we estimate for Satellites~1 and~2, reaching $M_*=1.3\times10^9~\msun$ at $z=1.7$. Their metallicity at that redshift is $\feh=-1.0$, which is expected to evolve to $\feh=\{-0.8,-0.5\}$ at $z=\{1,0.5\}$ \citep{forbes10}. These metallicities provide an excellent match to those of the old open clusters associated with the `Canis Major' GCs, which are $\feh=-0.45$ \citep{forbes04}. Together, Sagittarius and the `Canis Major' event represent the minority of GCs on the satellite branch, making it likely that the latter is the less massive of Satellites~1 and~2. We therefore associate the galaxy contributing the `Canis Major' GCs with Satellite~2.

Based on their ages ($\tau<12.5~\gyr$), metallicities ($\feh<2-0.25\tau/\gyr$, which gives $\feh<\{-0.5,-1.0\}$ at $\tau=\{10,12\}~\gyr$), and galactocentric radii ($R>5~\kpc$), there are 18 further GCs associated with the satellite branch of the GC age-metallicity distribution  in \autoref{fig:mwevo_zoom}.\footnote{A small number of GCs with ages $\tau>12.5~\gyr$ or galactocentric radii $R<5~\kpc$ may still have an ex-situ origin, but we deliberately adopt conservative sample limits.} These are NGC~362, NGC~1261, NGC~3201, NGC~5139, NGC~5272, NGC~5897, NGC~5904, NGC~5946, NGC~6121, NGC~6284, NGC~6544, NGC~6584, NGC~6752, NGC~6864, NGC~6934, NGC~6981, NGC~7006, and NGC~7089. While some of the oldest of these GCs may have formed in-situ, we expect a large fraction of them to have been accreted. Six of these GCs (NGC~362, NGC~3201, NGC~5139, NGC~6121, NGC~6934, NGC~7089) have retrograde orbits \citep[see][although both studies do not agree on each of these GCs individually]{dinescu99,allen06}, whereas the others do not. In general, these GCs have a variety of orbital kinematics. This suggests the existence of more than one progenitor (of which the retrograde one likely had NGC~5139/$\omega$~Cen as its nuclear cluster, cf.~\citealt{bekki03}), but the lack of structure in the satellite branch does not enable us to distinguish these.

In view of the above discussion, we refrain from associating individual GCs with Satellite~1, but encourage future studies to look for phase space correlations between the above set of GCs or those that have been proposed to be associated with Sagittarius or the `Canis Major' event (e.g.~using {\it Gaia}, \citealt{gaia16}). As such, the above GCs represent a `wish list' of interesting targets for kinematic follow-up work. This will enable identifying which of them belong to the inferred Milky Way progenitors. This may include GCs of which the age has not been measured to date, as the identified progenitors have sufficiently high masses to have hosted additional GCs. Such efforts may also identify groups of GCs originating from other accreted satellites, in addition to the three that are specifically identified in this work. After all, we estimate a total number of $\nbr=15.1\pm3.3$ accreted satellites, of which about half likely brought in GCs (see Section~\ref{sec:comp}).

Similarly, we identify six disc stars from the \citet{haywood13} sample in \autoref{fig:mwevo} that have metallicities $\feh<-0.65$ and ages $\tau<11~\gyr$, making them a natural extension of the GC satellite branch. If this interpretation is correct, it leads to the plausible conclusion that at least one of the satellites continued forming field stars after its youngest GC and accreted at $z<1$. If the satellite was accreted already at larger redshifts, the stars likely formed in the MW from the gas brought in by the satellite. In both cases, their origin is linked to satellite accretion. These stars are also shown in \autoref{fig:mwevo_zoom} and are all part of the {\it Hipparcos} catalogue, with identifiers HIP~36640, HIP~54641, HIP~57360, HIP~77637, HIP~80611, and HIP~102046. All six stars are part of the second {\it Gaia} data release (DR2, \citealt{gaia18}), listing high-precision positions, parallaxes, radial velocities, and proper motions. This provides the information necessary to verify if these stars are kinematically associated with the above candidate GCs and with (the remnant streams of) Sagittarius, the galaxy contributing the `Canis Major' GCs, or Satellite~1.

Satellite~1 remains unassociated with any known satellites of the Milky Way or their tidal debris. Therefore, it is likely to have been accreted several $\gyr$ ago. \citet[also see \citealt{pillepich14}]{deason13} suggest that the break in the density profile of the Galactic stellar halo at $R\sim25~\kpc$ \citep[e.g.][]{watkins09,sesar11} indicates an accretion event some 6--$9~\gyr$ ago (corresponding to $z=0.6$--$1.3$) of a single satellite that was considerably more massive than other accreted satellites. This finding is supported by independent evidence from halo kinematics \citep{belokurov18,deason18b}. In the context of \autoref{fig:mwevo}, it is natural to propose that this event corresponds to the accretion of Satellite~1, which would have had a mass of $M_*\sim2\times10^9~\msun$ at that time. Interestingly, the median galactocentric radius of the 18 candidate GCs associated with Satellite~1 listed above is $\sim8~\kpc$, which is considerably smaller than the nucleus of Sagittarius (NGC~6715, $\sim19~\kpc$, \citealt{harris96}) and also smaller than the putative nucleus of the galaxy contributing the `Canis Major' GCs (NGC~2808, $\sim11~\kpc$, \citealt{harris96}). This suggests that these GCs were brought in by a massive progenitor that spiralled in further due to dynamical friction than any other satellite. The results of this paper show that the echo of this satellites's existence is not only present in the density profile and kinematics of the stellar halo, but can also be seen in the age-metallicity distribution of Galactic GCs. Acknowledging the enigmatic nature of this satellite, which was plausibly the most massive satellite ever to have merged with the Milky Way, we propose to name it {\it Kraken}, after the elusive giant sea monsters that were rumoured to sink ships many centuries ago.

\subsection{The relative contributions of in-situ and ex-situ GCs} \label{sec:exsitu}

Following on from the above discussion, we can use the inferred assembly history of the Milky Way to determine the part of the GC population that formed in-situ and ex-situ. The Milky Way has 91 GCs with $M>10^5~\msun$ \citep[2010 edition, using $M_{V,\odot}=4.83$ and $M/L_V=2~\msun~{\rm L}_\odot^{-1}$]{harris96}. This agrees with the number of GCs expected to be present in the main progenitor at $z=0$ from \autoref{tab:prog} ($\ngcp=150_{-75}^{+12}$), but also shows that the Milky Way is at the low end of the expected range, possibly due to a lack of mergers with GC-rich galaxies. Out of these 91 GCs, 86~per~cent have $\feh<-0.5$, for which \autoref{fig:mwevo_zoom} shows that about 50~per~cent are associated with the satellite branch.\footnote{Based on sect.~5.1 of \citet{kruijssen18b}, this estimate includes half of the `ambiguous' GCs, i.e.~the crosses in \autoref{fig:mwevo_zoom} with $\feh<2-0.25\tau/\gyr$.} If we make the plausible assumption that the GCs at $\feh>-0.5$ formed in-situ, this implies that about 43~per~cent of the Galactic GCs have an ex-situ origin, i.e.~39 out of the 91 massive GCs come from accreted satellites, whereas the remaining 52 out of 91 formed in the main progenitor. In the absence of further evidence, it is reasonable to assume the same fractions apply to GCs with $M<10^5~\msun$. For a total population of 157 Galactic GCs in the \citet[2010 edition]{harris96} catalogue, we thus expect 67 with an ex-situ origin and predict that 90 formed in-situ.

Unsurprisingly, the fraction of GCs with an ex-situ origin depends on the metallicity. The ex-situ fraction must approach zero at high metallicities ($\feh>-0.5$), because the Milky Way has not accreted any satellites sufficiently massive to provide such metal-rich GCs (see Section~\ref{sec:disc} and e.g.~\citealt{mackey04}), but it increases towards low metallicities. We do not expect the ex-situ fraction to reach unity at any metallicity -- even at the lowest metallicities, a significant fraction of the metal-poor GCs should have formed in-situ during the early enrichment of the main progenitor of the Milky Way. For the \emosaics galaxies, about 50~per~cent of the GCs with metallicities $\feh<-1.5$ typically formed in-situ (see figs.~8 and~9 of \citealt{kruijssen18b}). For the Milky Way, \{25, 50\}~per~cent of the GCs with $\feh<-1.5$ are contained within a galactocentric radius of $R<\{5, 10\}~\kpc$, which suggests that many of these GCs also have an in-situ origin (compare figs.~2 and~8 of \citealt{kruijssen18b}). This conclusion differs from previous work, which often assumes an ex-situ origin for all metal-poor GCs not showing disc-like kinematics and chemistries \citep[e.g.][]{leaman13}.

\subsection{Comparison to other constraints on the formation and assembly of the Milky Way} \label{sec:comp}

Our constraints on the Milky Way's assembly and accretion history are consistent with a variety other studies, but we also find important differences relative to (and extensions of) previous results. The total stellar mass growth history integrated over all progenitors matches the observed star formation history of the Milky Way, with 50~per~cent of its stars predicted to have formed at $\tf=10.1\pm1.4~\gyr$ and the observed star formation history reaching this point at $\tf=10.5\pm1.5~\gyr$ \citep{snaith14}. The same applies to the inferred metal enrichment history of the main progenitor, which provides an excellent match to the age-metallicity distribution of the thick disc field stars \citep{haywood13} and to the enrichment history derived from the star formation history \citep{snaith15}.

Overall, our results imply that the Milky Way must have formed very rapidly in comparison to galaxies of similar masses. This conclusion stems from our comparisons to the suite of 25 \emosaics galaxies (see Section~\ref{sec:mwinf}), to the halo mass growth histories predicted by the extended Press-Schechter formalism \citep[][see Section~\ref{sec:mwinf}]{correa15b}, to stellar mass growth histories expected from abundance matching \citep[][see Section~\ref{sec:mwtree}]{behroozi13,moster13}, and to the observed stellar mass growth history of Milky Way progenitors across cosmic time \citep[see~\autoref{fig:mwevo}]{papovich15}. In addition, our conclusion of the rapid assembly of the Milky Way is supported by the findings of \citet{mackereth18}, who show that the presence of a bimodal $\afe$ distribution of the Galactic disc field stars is a rare feature among Milky Way-mass galaxies in \eagle, indicating an early phase of vigorous gas accretion and rapid star formation. These results all show that the formation and assembly history of the Milky Way is atypical for $L^\star$ galaxies.\footnote{An immediate implication of this result is that simulations of `typical' Milky Way-mass galaxies are unlikely to reproduce the properties of the Galactic GC population, unless the adopted GC formation and evolution model is missing critical physical ingredients.}

The agreement with previous findings also extends to the satellite progenitors. For instance, \citet{deason15,deason16} show that the accreted stellar mass of Milky Way-mass galaxies is mostly supplied by massive satellites, covering the stellar mass range $M_*=10^8$--$10^{10}~\msun$, or $M_*=10^8$--$10^9~\msun$ for galaxies with a relatively quiescent merger history such as the Milky Way. This mass range agrees very well with the expected masses of Satellites~1--3 prior to their accretion, which in Section~\ref{sec:known} are inferred to range from a few $10^8$ to $2\times10^9~\msun$. Given that these most massive satellites merged late (at $z<2$, see Section~\ref{sec:known}), our comparison of the satellite progenitor population ($M_*<2\times10^8~\msun$) with the stellar mass growth history of the main progenitor in Section~\ref{sec:mwtree} shows that the Milky Way did not experience any major mergers (i.e.~those with stellar mass ratios $>$1:4) since $z\sim4$ (corresponding to the past $\sim12.3~\gyr$). This extends previous literature constraints, which ruled out major mergers at $z\la2$ \citep[i.e.~the past $\sim10.5~\gyr$, see e.g.][]{wyse01,hammer07,stewart08,shen10,blandhawthorn16}. We also find that several of the satellites underwent (possibly major) mergers prior to their accretion onto the Milky Way, because the total number of progenitors ($\nleaf=24.1\pm10.2$) exceeds the number of mergers or accretion events experienced by the Milky Way ($\nbr=15.1\pm3.3$), even if the former carries a considerable uncertainty. It is not unusual for accreted satellites to have undergone previous merging -- \citet{deason14} predict that about one third of all dwarf galaxies within the virial radius experienced a major (defined in their study as having a stellar mass ratio $>1:10$, whereas we use $>1:4$) merger since $z\sim4$ \citep[also see e.g.][]{amorisco14}.

The inferred total number of $\sim15$ accretion events exceeds the number of seven accretion events of satellites hosting GCs obtained by \citet{mackey04}, illustrating that only about half of the Milky Way's progenitors hosted GCs. Indeed, \citet{mackey04} suggest that 6-11 progenitors may not have hosted any GCs, which implies a total of 13--18 mergers and exactly matches our estimate of $\nbr=15.1\pm3.3$.\footnote{In order to match the total mass of the stellar halo, \citet{mackey04} do assume lower satellite masses than those of our inferred progenitors, with a typical mass scale in logarithmic mass space of $10^{7.8}~\msun$ as opposed to our $10^{8.3}~\msun$. However, as we will see below, a substantial fraction of accretion events contributes to the Galactic (thick) disc and bulge, implying that such low satellite masses are not required.} If we adopt the observed relation between the total GC population mass and the host galaxy's halo mass \citep[e.g.][]{blakeslee97,spitler09,durrell14,harris17} and assume a typical GC mass of $M=2\times10^5~\msun$, this requires that about half of the accreted satellites had halo masses $\mvir\la10^{10}~\msun$ (which have $\ngc<1$), corresponding to stellar masses $M_*\la2\times10^7~\msun$ \citep[e.g.][]{moster13,behroozi13}. The remaining, massive progenitors span the mass range $M_*=2\times10^7$--$2\times10^9~\msun$, with the three most massive progenitors corresponding to Satellites~1--3 identified here, i.e.~Sagittarius, the galaxy contributing the `Canis Major' GCs, and Kraken. The prediction that Kraken, the most massive satellite, was accreted 6--$9~\gyr$ ago is also consistent with the idea that satellites surviving to $z=0$ are less massive than the previously disrupted ones \citep{sales07}. This emphasises the point that the GC population of the Milky Way's current satellite system is not necessarily representative for the complete ex-situ population.

With a typical mass scale of $M_*\sim2\times10^8~\msun$ and a range around this mass scale of two orders of magnitude ($M_*=2\times10^7$--$2\times10^9~\msun$), the inferred GC-bearing progenitors cover a narrower mass range and have higher masses than proposed by \citet{leaman13} based on the Galactic GC age-metallicity distribution. These authors find masses in the range $M_*=3\times10^5$--$10^9~\msun$, with a typical mass scale of $M_*\sim3\times10^7~\msun$. The difference is largely caused by the fact that \citet{leaman13} associate all old, metal-poor GCs with (predominantly low-mass) satellites, whereas we consider a large fraction of these to have formed in the precursors of the Milky Way's main progenitor and its most massive accreted satellites. The latter interpretation is more consistent with the other observational constraints on the formation and assembly of the Milky Way discussed above. An additional source of discrepancy may be that there exists some tension (of $\Delta\feh\sim0.3$) between the observed galaxy mass-metallicity relation and that produced by the \eagle Recal-L025N0752 simulation at galaxy masses $M_*<10^9~\msun$ \citep[fig.~13]{schaye15}. This is not necessarily an intrinsic bias of \eagle, as nucleosynthetic yields and metallicity calibrations are uncertain at a similar level \citep{schaye15}. Irrespective of the reason, if the simulations systematically overestimate the metallicities of low-mass galaxies, then this would lead to a corresponding overprediction by a factor of several of the satellite progenitor masses reported here. The work by \citet{leaman13} uses empirical age-metallicity relations of satellite galaxies in the Local Group to estimate satellite progenitor masses. These carry an uncertainty of a similar magnitude due to the choice of metallicity calibration, but this is independent from any bias possibly present in \eagle.

Following from the inferred formation and assembly history of the Milky Way, we estimate that 67 out of the 157 Galactic GCs from the \citet[2010 edition]{harris96} catalogue (i.e.~a little more than 40~per~cent) have an ex-situ origin. This is higher than the number derived by \citet[40 ex-situ GCs]{mackey04} based on a comparison of the core radii of Galactic GCs to these of GCs in dwarf galaxies. However, the difference is largely caused by our inclusion of about half of the old, metal-poor GCs in the part of the GC age-metallicity distribution where the satellite branch is indistinguishable from the main branch. This choice is motivated by the ex-situ fractions obtained for this part of age-metallicity space in the \emosaics galaxies \citep{kruijssen18b} and is not necessarily inconsistent with the idea from \citet{mackey04} that inflated GC core radii trace the ex-situ origin of GCs -- if the natal dwarf galaxies of the oldest GCs were accreted by the Milky Way during its initial collapse at $z>4$, then the properties of these satellites and the Milky Way's main progenitor at that time may have been similar, resulting in similar core radii. As such, the estimated number of Galactic GCs formed ex-situ from \citet{mackey04} is a lower limit.

By combining the observed fraction of dwarf galaxies hosting nuclear clusters with the satellite progenitor statistics derived here, we can also make an estimate of the number of former nuclear clusters among the Galactic GC population. Across the galaxy mass range of the GC-bearing progenitors considered here ($M_*=2\times10^7$--$2\times10^9~\msun$), a fraction $f_N\approx0.65$ of dwarf galaxies in the Fornax Cluster host nuclear clusters \citep{munoz15,ordenesbriceno18}, whereas $f_N\approx0.15$ for the remaining progenitors ($M_*=4.5\times10^6$--$2\times10^7~\msun$). If the same applies to the progenitors of the Milky Way and all nuclear clusters survived the accretion and tidal disruption of their host, then the roughly equal estimated number of progenitors among these two mass bins implies a mean nucleation fraction of $f_N\approx0.4$ across the $\nbr=15.1\pm3.3$ progenitors of the Milky Way. Based on these numbers, we predict that the Galactic GC population hosts $N_{\rm nucl}=6.0\pm1.3$ former nuclear clusters, which is similar to the number of Galactic GCs for which this has been proposed in the observational literature (see e.g.~sect.~4.1 of \citealt{pfeffer14} for a discussion). These `GCs' are likely to exhibit spreads in $\feh$ and possibly age.

Finally, the satellite progenitors of the Milky Way identified in this paper imply a total accreted mass of $M_*\sim3$--$4\times10^9~\msun$. Naively, this may seem to violate the (smaller) total mass of the Galactic stellar halo \citep[$10^9~\msun$, e.g.][]{freeman02}. However, it is likely these masses are consistent, because not all of the accreted mass has been deposited in the halo. For instance, the galaxy contributing the `Canis Major' GCs ($\sim10^9~\msun$) may largely have deposited stars in the thick disc \citep{martin04}, whereas the central spheroids of other dwarfs are likely to survive tidal stripping and eventually spiral in due to dynamical friction, thus merging with the Galactic disc or bulge and depositing a large fraction of their stellar mass there \citep[e.g.][]{meza05,amorisco17}.

\subsection{Dependence of results on the absolute GC age calibration} \label{sec:cal}

The main caveat of the presented analysis is posed by the large uncertainties on the age measurements of GCs. This does not as much affect their relative age differences,\footnote{This  is illustrated by the considerable structure in age-metallicity space (see e.g.~\autoref{fig:agezmw}), which would have been washed out if the relative ages were uncertain.} but the absolute (or `reference') GC age hinges critically on the GC age calibration. Due to these uncertainties, many observational studies focus on relative ages \citep[e.g.][]{marinfranch09,wagnerkaiser17b}. Given that the ages of the Galactic GCs lean towards the highest redshifts at which stars of the corresponding metallicities are formed in the $25~\mpc$ \eagle Recal-L025N0752 volume (see \autoref{fig:agezmwmatch}), we expect that any systematic offset in the absolute ages is one that would have caused them to be overestimated.

If the absolute ages of Galactic GCs were smaller than listed in the three adopted literature catalogues \citep{forbes10,dotter10,dotter11,vandenberg13} and our combined mean sample (Appendix~\ref{sec:appobs}), then this has several implications for the derived formation and assembly history of the Milky Way. Firstly (and trivially), the Galaxy would not have undergone as rapid mass growth and metal enrichment as the data currently require. Secondly, the quantities derived from the median age $\widetilde{\tau}$ and the GC metal enrichment rate $\dfehdt$ (see \autoref{tab:fitmw}) would be biased towards lower values (they all correlate positively with these two age-based metrics). Thirdly, the number of GCs per halo derived from the bottom panel of \autoref{fig:agezmwmatch} would decrease, such that the Galactic GC population would be assembled from a larger number of progenitors. In general, each of these implications of a shift towards younger ages results in a more gradual formation and assembly of the Milky Way, with a richer merger tree and $\gg3$ satellite progenitors that should be associated with the satellite branch in the GC age-metallicity distribution.

Importantly, a downward revision of the observed GC ages would lead to a number of problems. Firstly, many of the galaxy-related quantities in \autoref{tab:fitmw} are obtained from more than one GC-related metric, which provides a useful consistency check. Quantities derived from absolute GC ages ($\widetilde{\tau}$ and $\dfehdt$) are currently consistent with those that are estimated using other, independent metrics. Secondly, it would be hard to reproduce the well-defined nature of the satellite branch in \autoref{fig:agezmwmatch} with a large number of satellite progenitors. Especially at the (even lower) required masses, the progenitors would be expected to form stars at a considerably wider range of metallicities at a given redshift. Otherwise, it would be highly coincidental that all of these progenitors had identical metal enrichment histories. Finally, the results obtained from our GC-based analysis are found to be consistent with independent measurements from the literature, such as the star formation history, metal enrichment history, and field star ages and metallicities, both for the Galactic disc and for the `Canis Major' and Sagittarius dwarfs. Together, these considerations argue against a significant systematic bias of the GC ages.

Finally, we note that our results are insensitive to differences in the absolute GC age calibration {\it between} the three literature samples used. For the subset of 36 GCs with masses $M>10^5~\msun$ and metallicities $-2.5<\feh<-0.5$ that appear in each of these samples, the median ages from the \citet{forbes10}, \citet{dotter10,dotter11}, and \citet{vandenberg13} catalogues are $\overline{\tau}=\{12.26, 12.75, 12.01\}~\gyr$, respectively. These mean ages agree to within the age uncertainties listed in Appendix~\ref{sec:appobs}, showing that the adopted uncertainties already reflect any possible biases. If we change the median age $\widetilde{\tau}$ of the Galactic GCs by a nominal $\Delta\tau=0.5~\gyr$, each of the derived quantities describing the Milky Way's formation and assembly history in \autoref{tab:fitmw} changes by less than the quoted uncertainty. This insensitivity to any biases in GC age arises, because the relations used for inferring these quantities exhibit a scatter that is similar in magnitude to the slope of the relation in units of $\gyr^{-1}$ \citep[see table~7 of][]{kruijssen18b}. This means that a systematic age bias in excess of $1~\gyr$ is needed to significantly change the results. The same applies to the other quantities that could be sensitive to the absolute age calibration, such as $\dfehdt$. Because the GC ages that have been measured since 2001 agree to within 5--10~per~cent \citep[i.e.~$<1~\gyr$;][]{dotter13}, we conclude that our results are robust against systematic age uncertainties, unless all GC age measurements share the same systematic bias.

Further improvement of the GC age uncertainties may be achieved using highly accurate distances to reference stars that can be obtained with {\it Gaia} \citep{gaia16,gaia16b}. However, the measurements of chemical abundances will remain a major uncertainty for the foreseeable future. There exist systematic differences in metallicity and abundance scales adopted by various studies reporting GC ages \citep[e.g.][]{carretta97,kraft03,carretta09b}, which in turn affect the GC age measurements \citep[e.g.][]{vandenberg16}.\footnote{As shown by \citet{vandenberg13}, uncertainties in the metallicity scale mainly affect the ages derived for metal-rich GCs ($\feh>-0.7$) by \citet{marinfranch09}. Our conclusions do not change when accounting for this source of uncertainty, because we omit GCs with $\feh>-0.5$ and average over three different GC age-metallicity samples (of which the ages from \citealt{marinfranch09} is contained in the \citealt{forbes10} compilation).} These may be improved by better measurements of the stellar mass-luminosity relation, which is constrained by studies of eclipsing binaries \citep[e.g.][]{henry93,malkov07}. The main obstacle for applying these constraints to GCs is that they exhibit internal spreads in helium abundance \citep[e.g.][]{bastian18}, which in turn modifies the stellar mass-luminosity relation \citep[e.g.][]{salaris06,vandenberg14,chantereau16}. Given the major importance of accurate GC ages for reconstructing the formation and assembly history of the Milky Way and potentially other galaxies, we stress the importance of continued work on these issues.

\subsection{Dependence of results on the adopted GC model} \label{sec:gcmod}

An additional caveat may come from the physical model used to calculate the formation and disruption of the GCs in \emosaics. Specifically, we use physical prescriptions for the cluster formation efficiency and the (high-mass end of the) initial cluster mass function, as well as for GC mass loss due to the local tidal field that they experience in the simulations. If these prescriptions are incorrect, either in terms of their absolute values or their environmental scaling, then this may bias the GC age-metallicity distribution.

The fundamental approach of \emosaics is to investigate whether the observed properties of GC populations can be reproduced when assuming that they formed according to local-Universe physics. The prescriptions for the cluster formation efficiency, initial cluster mass function, and cluster disruption, as well as their environmental dependence have all been extensively tested against observations of stellar cluster populations in nearby galaxies, to the point that they accurately describe the observations to within $<0.3~\dex$. In \citet{pfeffer18}, \citet{reinacampos18b}, \citet{usher18}, and Pfeffer et al.~in prep., we show that switching off the environmental dependences of the cluster formation efficiency and the (high-mass end of the) initial cluster mass function results in young cluster populations and GC populations that are inconsistent with observations in terms of their spatial distributions, formation histories, mass-metallicity distributions, luminosity functions, and age distributions. In addition, the restriction of the analysed GC sample to those with $M>10^5~\msun$ minimises any influence of GC disruption, which mostly affects GCs of lower masses \citep[also see][]{kruijssen18b,reinacampos18}. These points suggest that the uncertainties introduced by the adopted GC formation and disruption model are minor compared to the age uncertainties discussed in Section~\ref{sec:cal}.

\section{Summary and implications} \label{sec:concl}

In this paper, we have reconstructed the formation and assembly history of the Milky Way using the age-metallicity distribution of its GC population. To achieve this, we have used the results of the \emosaics project, which is a volume-limited suite of 25 cosmological hydrodynamical zoom-in simulations of Milky Way-mass galaxies that includes a self-consistent model for the formation and disruption of stellar cluster populations \citep{pfeffer18,kruijssen18b}, and applied these to a compilation of ages and metallicities of 96 Galactic GCs (see Appendix~\ref{sec:appobs}). Using the GC age-metallicity distribution and its correlation with the host galaxy's formation and assembly history \citep[see][]{kruijssen18b}, we derive 12 different quantities describing the Milky Way's mass growth history, its metal enrichment history, and its merger tree. These results show that, in terms of its integrated properties, the Milky Way falls close to the median of the \emosaics galaxies. However, in terms of quantities describing its growth, assembly, and metal enrichment rate, the Milky Way is highly unusual, spanning the 72th-94th percentile of the \emosaics simulations. This shows that the Milky Way formed extremely rapidly, which we quantify by comparing our results to the halo mass growth histories predicted by the extended Press-Schechter formalism \citep{correa15b}, to the stellar mass growth histories expected from abundance matching \citep{behroozi13,moster13}, and to the observed stellar mass growth history of Milky Way progenitors across cosmic history \citep{papovich15}.

In addition to these results based on the statistical correlation between the GC age-metallicity distribution and galaxy formation and assembly, we also use the detailed distribution of GCs in age-metallicity space to reconstruct the merger tree of the Milky Way. This is done by comparison to the median properties of galaxies in the $25~\mpc$ \eagle Recal-L025N0752 volume \citep{schaye15} at a given redshift and metallicity of newly-formed stars. We find that GC metallicity is the strongest indicator of the mass of its host galaxy at formation, with a secondary dependence on GC age. Using this technique, we identify the individual progenitors of the Milky Way that contributed GCs, as well as their stellar mass growth and metal enrichment histories over the age range spanned by their GCs. The results obtained this way are shown to be consistent with and extend the previous observational constraints of the Milky Way's metal enrichment history (see \autoref{fig:mwevo}). Using the number of GCs expected to be contributed by the accreted satellites, we estimate that $\sim40$~per~cent of the Galactic GCs have an ex-situ origin. Out of the 157 GCs in the \citet[2010 edition]{harris96} catalogue, this corresponds to 60--70 GCs, with the remaining $\sim90$ having formed in-situ.

\begin{figure*}
\includegraphics[width=0.792\hsize]{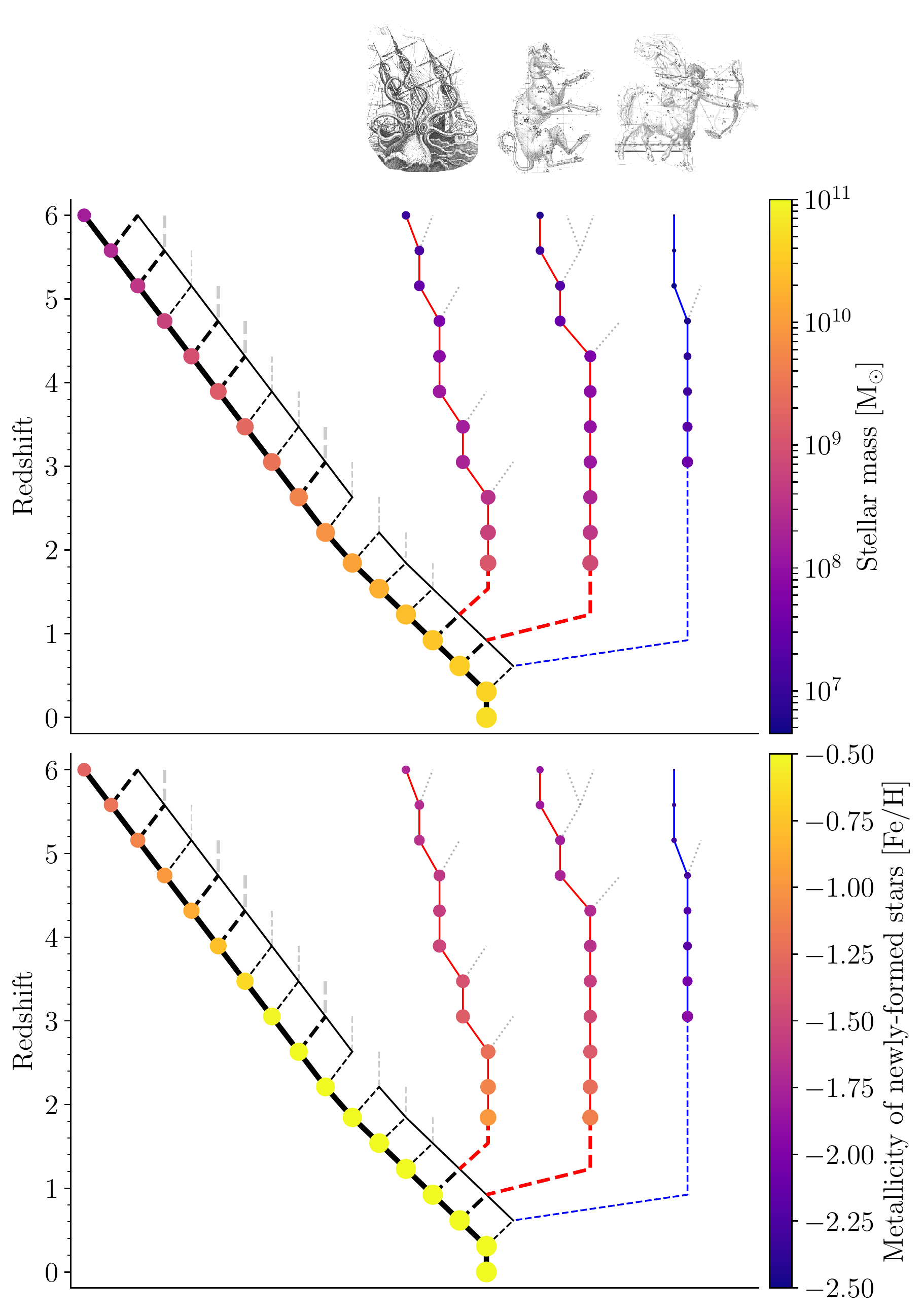}%
\caption{
\label{fig:mwtree}
Galaxy merger tree of the Milky Way, inferred using the Galactic GC age-metallicity distribution, colour-coded by galaxy stellar mass (top panel) and the metallicity of newly-formed stars (bottom panel). This figure summarises many of the results presented in this paper. The main progenitor is denoted by the thick black line, whereas the coloured lines indicate the identified (and likely most massive) satellites, i.e.~red lines refer to Satellite~1 (left) and Satellite~2 (right), whereas the blue line shows Satellite~3. Dashed lines indicate mergers that are topologically robust, but have unknown timing within a given redshift interval. Thin lines mark tiny mergers (with mass ratio $<$1:100), whereas lines of intermediate thickness denote minor (or possibly major) mergers (with mass ratio $>$1:100). Light grey lines represent progenitors that are included to illustrate the inferred global statistics of the Milky Way's merger tree, but have no absolute meaning. These merger trees are consistent with the total number of mergers ($\nbr$), the number of high-redshift mergers ($\nbrz$), the total number of progenitors ($\nleaf$), and the numbers of tiny and minor mergers ($N_{<1:100}$ and $N_{1:100-1:4}$) from \autoref{tab:fitmw}, as well as with the identified progenitors, their connection to known satellites from Section~\ref{sec:known}, and their evolution from \autoref{tab:prog}. Note that only progenitors with masses $M_*>4.5\times10^6~\msun$ are included. From left to right, the three images along the top of the figure indicate the three identified dwarf galaxies Kraken, the progenitor contributing the `Canis Major' GCs, and Sagittarius.
}
\end{figure*}
\autoref{fig:mwtree} summarises many of the results presented in this paper by visualising them in terms of the inferred merger tree of the Milky Way. This merger tree captures the following quantitative results from Section~\ref{sec:recon} (see specifically \autoref{tab:fitmw} and \autoref{tab:prog}), as well as the interpretation from Section~\ref{sec:disc}.
\begin{enumerate}
\item
The main progenitor of the Milky Way experienced a total of $\nbr=15\pm3$ mergers (or `branches' of the merger tree), with $\nbrz=9\pm2$ of these taking place at $z>2$.
\item
Out of these mergers, $N_{<1:100}=8\pm2$ represent `tiny' mergers with mass ratios $<$1:100, whereas the remainder represent minor (mass ratio 1:100--1:4) or possibly major (mass ratio $>$1:4) mergers. The Milky Way experienced no major mergers since $z\sim4$.
\item
The total number of progenitors is $\nleaf=24\pm10$ (or `leaves' of the merger tree), implying that (after subtraction of the Milky Way's main progenitor and its direct mergers) about 8 progenitors must have merged with other progenitors prior to their accretion onto the Milky Way's main progenitor.
\item
Through a combination of vigorous in-situ star formation and satellite accretion, the main progenitor grows rapidly, attaining 10~per~cent of its final mass at $z=2.5$, and 50~per~cent at $z=1.2$.
\item
The GC age-metallicity distribution specifically indicates the presence of three satellite progenitors. Because these hosted most of the ex-situ GCs, these are likely the most massive satellites to have been accreted by the Milky Way.
\item
Out of the three identified satellites, the least massive one reaches a mass of $M_*\sim4\times10^7~\msun$ at $z=3$. It likely corresponds to the Sagittarius dwarf galaxy, with a predicted stellar mass prior to accretion of a few $10^8~\msun$, consistent with previous estimates. This low mass implies that it was accreted by the Milky Way through a tiny merger.
\item
The remaining two identified satellites are indistinguishable on the satellite branch in the GC age-metallicity distribution, but (at least) two such progenitors are required to explain the large number of GCs on the satellite branch. They reach a mass of $M_*\sim1\times10^9~\msun$ at $z=1.7$ and thus must be part of the $\sim6$ galaxies that are accreted at $z<2$. Given their relatively high masses, these galaxies were accreted as minor mergers rather than tiny ones. One of these two massive satellite progenitors is suggested to have brought in the GCs historically associated with the Canis Major `dwarf' (see footnote~\ref{ft:cma}), with a mass of $M_*=1$--$2\times10^9~\msun$ at the time of accretion.
\item
The presence of many (relatively metal-rich) GCs on the satellite branch unassociated with any known satellites predicts the existence of a hitherto unknown progenitor that likely represents the most massive satellite ever to have been accreted by the Milky Way, with a mass of $M_*\sim2\times10^9~\msun$ at the time of accretion. We name this enigmatic galaxy {\it Kraken} and propose it is responsible for the break in the density profile of the stellar halo at $R\sim25~\kpc$ \citep[e.g.][]{watkins09,sesar11}, which requires it to have been accreted between $z=0.6$--$1.3$ \citep{deason13} and implies that its tidal debris has since dispersed.
\end{enumerate}
These results and predictions are highly suitable for observational tests using the 6D phase space information of the Galactic stellar halo and the GC population. With the arrival of {\it Gaia}, it is possible to perform a targeted search for the stars and GCs that are associated with the identified satellite progenitors of the Milky Way. For instance, Section~\ref{sec:known} lists the GCs that may be associated with Kraken based on their ages and metallicities. This set of candidates can be pruned using their 3D positions and kinematics, thus providing the GCs that formed in Kraken with high confidence. More broadly, the statistical properties of the Milky Way's progenitors summarised above and in \autoref{fig:mwtree} are testable in future studies of stellar abundances and their phase space information.\footnote{This paper was submitted prior to {\it Gaia} DR2. The analysis presented in this work, including the `wish lists' of interesting candidate GCs to be targeted by kinematic follow-up studies, did not change during the refereeing process. As such, the presented results are genuine predictions derived from the age-metallicity distribution of the Galactic GC population. After the submission of this paper and the release of {\it Gaia} DR2, \citet{myeong18} submitted a letter based on the DR2 catalogue, reporting a set of eight GCs with kinematics indicating a common origin (e.g.~orbital eccentricities $e>0.86$) and suggest that these may be responsible for the break in the stellar halo density profile identified by \citet{deason13}. Could these GCs have formed in our Satellites~1 or~2, which we refer to as Kraken and `Canis Major'? This is an exciting possibility, as the GCs identified by \citet{myeong18} indeed follow the satellite branch in age-metallicity space traced by these two progenitors. The \citet{myeong18} GC sample closely matches the `Canis Major' candidate GCs listed in Section~\ref{sec:known}, implying that these were likely not brought in by Kraken. Indeed, the apocentre radii of the \citet{myeong18} GCs ($R=12$--$20~\kpc$) are smaller than the break radius of the stellar halo ($R\sim25~\kpc$), which we suggest was sculpted by the accretion of Kraken onto the Milky Way. If this feature was instead generated by the `Canis Major' GCs as suggested by \citet{myeong18}, it would require the GCs to be deposited onto orbits with apocentre radii smaller than the break radius. This question aside, the good match between the `Canis Major' candidates from Section~\ref{sec:known} and the \citet{myeong18} sample is encouraging, as it demonstrates that the complementary approaches taken by both studies can lead to consistent results. The interpretation of these findings and the association of specific Galactic GCs with the satellite progenitors identified in this work remain important open questions, which we expect will be answered through a systematic analysis of the {\it Gaia} DR2 data.}

Our constraints on the formation and assembly history of the Milky Way match and extend a broad range of previous findings from the literature, but also differ in some respects. For instance, the presented mass growth history and metal enrichment history of the Milky Way's main progenitor agree with (and add to) earlier results based on the field star population of the thick disc \citep{haywood13,snaith14,snaith15}. Our conclusion that the Milky Way did not experience any major mergers since $z\sim4$ sharpens the previous upper limit of having had no major mergers since $z\sim2$ \citep{wyse01,hammer07,stewart08}. Turning to the Milky Way's satellite progenitors, the mass range of the three identified, most massive accreted satellites ($M_*=2\times10^8$--$2\times10^9~\msun$) is consistent with the upper end of the satellite progenitor mass range of Milky Way-mass galaxies with relatively quiescent assembly histories \citep{deason16}. The total number of $15\pm3$ mergers is consistent with previous estimates based on the structural properties of GCs \citep[13--18 mergers, see][]{mackey04}, even if the satellite masses are higher by a factor of several in our interpretation. This difference arises because we do not assume that all of their stellar mass is deposited into the stellar halo, but expect a significant fraction to be added to the Galactic disc and bulge. About half of the accreted satellites had masses too low to have been associated with GCs ($M_*\la2\times10^7~\msun$), implying that the remaining population spanning $M_*=2\times10^7$--$2\times10^9~\msun$ contributed the ex-situ GC population of the Milky Way. These masses are considerably higher and span a narrower range than the previous estimate of \citet{leaman13}, which results from their assumption that the entire old, metal-poor (halo) GC population formed ex-situ. As discussed in \citet{kruijssen18b}, the \emosaics simulations show that only about half of these GCs have an ex-situ origin, and often formed in the (at that time still low-mass) precursors to the most massive accreted satellites. Finally, we combine these results with the dwarf galaxy nucleation fractions observed in the Fornax Cluster and predict that the Galactic GC population hosts $N_{\rm nucl}=6\pm1$ former nuclear clusters, which are likely to exhibit a spread in $\feh$.

This work demonstrates that the Galactic GC population provides detailed insight into the formation and assembly history of the Milky Way. After a rapid phase of in-situ growth and copious dwarf galaxy accretion that lasted until $z=2$--$4$, the Milky Way evolved quiescently, accreting satellites of low-to-intermediate masses through tiny and minor mergers. The detailed information encoded in the GC age-metallicity distribution enables us to identify several of the most massive progenitors of the Milky Way and constrain their mass growth and metal enrichment histories, including the Sagittarius dwarf, the galaxy contributing the `Canis Major' GCs, and a newly postulated, most massive accreted satellite of the Milky Way, which we name {\it Kraken}. While the GC age-metallicity distribution is unlikely to provide meaningful insight into the formation of galaxies outside the virial radius of the Milky Way due to the poor precision of absolute GC ages at such distances, it may be possible to identify the youngest GCs in other galaxies within (or just outside) the Local Group and use these to infer the properties of their satellite progenitors. In addition, we expect GC-related observables other than their ages to facilitate comparable analyses of more distant galaxies. Examples of specific observables of interest are the spatial distribution of GCs, their metallicity distribution, their kinematics, and their association with stellar streams. In future work, we will explore these and other connections between GC populations and their host galaxies' formation and assembly histories.

\section*{Acknowledgements}

JMDK gratefully acknowledges funding from the German Research Foundation (DFG) in the form of an Emmy Noether Research Group (grant number KR4801/1-1, PI Kruijssen). JMDK and MRC gratefully acknowledge funding from the European Research Council (ERC) under the European Union's Horizon 2020 research and innovation programme via the ERC Starting Grant MUSTANG (grant agreement number 714907, PI Kruijssen). JP and NB gratefully acknowledge funding from the ERC under the European Union's Horizon 2020 research and innovation programme via the ERC Consolidator Grant Multi-Pop (grant agreement number 646928, PI Bastian). MRC is supported by a Fellowship from the International Max Planck Research School for Astronomy and Cosmic Physics at the University of Heidelberg (IMPRS-HD). NB and RAC are Royal Society University Research Fellows. This work used the DiRAC Data Centric system at Durham University, operated by the Institute for Computational Cosmology on behalf of the STFC DiRAC HPC Facility (\url{www.dirac.ac.uk}). This equipment was funded by BIS National E-infrastructure capital grant ST/K00042X/1, STFC capital grants ST/H008519/1 and ST/K00087X/1, STFC DiRAC Operations grant ST/K003267/1 and Durham University. DiRAC is part of the National E-Infrastructure. This study also made use of high performance computing facilities at Liverpool John Moores University, partly funded by the Royal Society and LJMUÕs Faculty of Engineering and Technology. The image of Kraken in \autoref{fig:mwtree} was adapted from an illustration in Pierre Denys de Montfort's `Histoire naturelle, g\'{e}n\'{e}rale et particuliere des Mollusques', Vol.~2, 1801, pp.~256--257. The images of Canis Major and Sagittarius in \autoref{fig:mwtree} were adapted from Alexander Jamieson's `A Celestial Atlas', 1822, plates~20 and~25. This work has made use of Matplotlib \citep{hunter07}, Numpy \citep{vanderwalt11}, Scipy \citep{jones01}, and Astropy \citep{astropy13}. We thank Alis Deason, Aaron Dotter, Wyn Evans, Duncan Forbes, Bill Harris, Ryan Leaman, and Don Vandenberg for helpful feedback on the original version of this paper, as well as Jacob Ward, Pieter van Dokkum, M\'{e}lanie Chevance, and Chris Usher for helpful discussions, and Misha Haywood for providing the stellar ages and metallicities from \citet{haywood13} in electronic form. We thank an anonymous referee for a helpful report that improved the presentation of this work.

\bibliographystyle{mnras}
\bibliography{mybib}

\appendix

\section{Observed properties of the GC population in the Milky Way} \label{sec:appobs}

We list the observed properties of the Galactic GC population in \autoref{tab:gcobs}. This includes the absolute $V$-band magnitudes from \citet[2010 edition]{harris96}, cluster masses assuming $M_{V,\odot}=4.83$ and $M/L_V=2~\msun~{\rm L}_\odot^{-1}$ \citep[e.g.][]{mclaughlin05,kruijssen09,strader11}, and the ages $\tau$ and metallicities $\feh$ from three literature GC samples \citep{forbes10,dotter10,dotter11,vandenberg13}, as well as a combined sample, which takes the mean whenever multiple measurements are available per GC. For all samples, the maximum age is set to $\tau_{\rm max}=\thub-0.18~\gyr$, which corresponds to $z=20$. The individual age uncertainties are listed for each GC, but we assume a universal metallicity uncertainty of $\sigma(\feh)=0.1~\dex$. For the \citet{vandenberg13} sample, the uncertainties include the statistical uncertainty listed in their paper, as well as an additional uncertainty of $1~\gyr$ to account for uncertainties in distance and chemical abundances. The error bars quoted for the other samples already include these sources of uncertainty. See Section~\ref{sec:mwobs} for further discussion.

\begin{table*}
  \caption{Observed properties of 96 Galactic GCs based on three main GC age-metallicity samples from the literature. From left to right, the columns show: GC name; absolute $V$-band magnitude $M_V$ from \citet[2010 edition]{harris96}, implied mass $M$ assuming $M_{V,\odot}=4.83$ and $M/L_V=2$; ages $\tau$ and metallicities $\feh$ from the \citet{forbes10}, \citet{dotter10,dotter11}, and \citet{vandenberg13} samples; mean age across all three samples $\overline{\tau}$; mean metallicity across all three samples $\overline{\feh}$.} 
\label{tab:gcobs}
  \begin{tabular}{l c c c c c c c c c c c c c}
   \hline
 &  &  & \multicolumn{2}{c}{\citet{forbes10}} & \rule{-7pt}{0ex} & \multicolumn{2}{c}{\citet{dotter10,dotter11}} & \rule{-7pt}{0ex} & \multicolumn{2}{c}{\citet{vandenberg13}} & \rule{-7pt}{0ex} & \multicolumn{2}{c}{Mean} \vspace{1mm} \\ \cline{4-5} \cline{7-8} \cline{10-11} \cline{13-14} \rule{-2pt}{3ex} 
   Name & $M_V$ & $\log M$ & $\tau$ & $\feh$ & \rule{-7pt}{0ex} & $\tau$ & $\feh$ & \rule{-7pt}{0ex} & $\tau$ & $\feh$ & \rule{-7pt}{0ex} & $\overline{\tau}$ & $\overline{\feh}$ \\ 
    & $[{\rm mag}]$ & $[\msun]$ & $[\gyr]$ &  & \rule{-7pt}{0ex} & $[\gyr]$ &  & \rule{-7pt}{0ex} & $[\gyr]$ &  & \rule{-7pt}{0ex} & $[\gyr]$ &  \\ 
   \hline
   NGC104 & $-9.42$ & $6.00$ & $13.06 \pm 0.90$ & $-0.78$ & \rule{-7pt}{0ex} & $12.75 \pm 0.50$ & $-0.70$ & \rule{-7pt}{0ex} & $11.75 \pm 1.03$ & $-0.76$ & \rule{-7pt}{0ex} & $12.52 \pm 0.49$ & $-0.75$ \\ 
   NGC288 & $-6.75$ & $4.93$ & $10.62 \pm 0.51$ & $-1.14$ & \rule{-7pt}{0ex} & $12.50 \pm 0.50$ & $-1.40$ & \rule{-7pt}{0ex} & $11.50 \pm 1.07$ & $-1.32$ & \rule{-7pt}{0ex} & $11.54 \pm 0.43$ & $-1.29$ \\ 
   NGC362 & $-8.43$ & $5.61$ & $10.37 \pm 0.51$ & $-1.09$ & \rule{-7pt}{0ex} & $11.50 \pm 0.50$ & $-1.30$ & \rule{-7pt}{0ex} & $10.75 \pm 1.03$ & $-1.30$ & \rule{-7pt}{0ex} & $10.87 \pm 0.42$ & $-1.23$ \\ 
   NGC1261 & $-7.80$ & $5.35$ & $10.24 \pm 0.51$ & $-1.08$ & \rule{-7pt}{0ex} & $11.50 \pm 0.50$ & $-1.35$ & \rule{-7pt}{0ex} & $10.75 \pm 1.03$ & $-1.27$ & \rule{-7pt}{0ex} & $10.83 \pm 0.42$ & $-1.23$ \\ 
   NGC1851 & $-8.33$ & $5.57$ & $9.98 \pm 0.51$ & $-1.03$ & \rule{-7pt}{0ex} & -- & -- & \rule{-7pt}{0ex} & $11.00 \pm 1.03$ & $-1.18$ & \rule{-7pt}{0ex} & $10.49 \pm 0.58$ & $-1.10$ \\ 
   NGC1904 & $-7.86$ & $5.38$ & $11.14 \pm 0.90$ & $-1.37$ & \rule{-7pt}{0ex} & -- & -- & \rule{-7pt}{0ex} & -- & -- & \rule{-7pt}{0ex} & $11.14 \pm 0.90$ & $-1.37$ \\ 
   NGC2298 & $-6.31$ & $4.76$ & $12.67 \pm 0.64$ & $-1.71$ & \rule{-7pt}{0ex} & $13.00 \pm 1.00$ & $-1.90$ & \rule{-7pt}{0ex} & -- & -- & \rule{-7pt}{0ex} & $12.84 \pm 0.59$ & $-1.80$ \\ 
   NGC2419 & $-9.42$ & $6.00$ & $12.30 \pm 1.00$ & $-2.14$ & \rule{-7pt}{0ex} & $13.00 \pm 1.00$ & $-2.00$ & \rule{-7pt}{0ex} & -- & -- & \rule{-7pt}{0ex} & $12.65 \pm 0.71$ & $-2.07$ \\ 
   NGC2808 & $-9.39$ & $5.99$ & $10.80 \pm 0.38$ & $-1.11$ & \rule{-7pt}{0ex} & -- & -- & \rule{-7pt}{0ex} & $11.00 \pm 1.07$ & $-1.18$ & \rule{-7pt}{0ex} & $10.90 \pm 0.57$ & $-1.14$ \\ 
   NGC3201 & $-7.45$ & $5.21$ & $10.24 \pm 0.38$ & $-1.24$ & \rule{-7pt}{0ex} & $12.00 \pm 0.75$ & $-1.50$ & \rule{-7pt}{0ex} & $11.50 \pm 1.07$ & $-1.51$ & \rule{-7pt}{0ex} & $11.25 \pm 0.45$ & $-1.42$ \\ 
   NGC4147 & $-6.17$ & $4.70$ & $11.39 \pm 0.51$ & $-1.50$ & \rule{-7pt}{0ex} & $12.75 \pm 0.75$ & $-1.70$ & \rule{-7pt}{0ex} & $12.25 \pm 1.03$ & $-1.78$ & \rule{-7pt}{0ex} & $12.13 \pm 0.46$ & $-1.66$ \\ 
   NGC4372 & $-7.79$ & $5.35$ & $12.54 \pm 0.90$ & $-1.88$ & \rule{-7pt}{0ex} & -- & -- & \rule{-7pt}{0ex} & -- & -- & \rule{-7pt}{0ex} & $12.54 \pm 0.90$ & $-1.88$ \\ 
   NGC4590 & $-7.37$ & $5.18$ & $11.52 \pm 0.51$ & $-2.00$ & \rule{-7pt}{0ex} & $13.00 \pm 1.00$ & $-2.30$ & \rule{-7pt}{0ex} & $12.00 \pm 1.03$ & $-2.27$ & \rule{-7pt}{0ex} & $12.17 \pm 0.51$ & $-2.19$ \\ 
   NGC4833 & $-8.17$ & $5.50$ & $12.54 \pm 0.64$ & $-1.71$ & \rule{-7pt}{0ex} & $13.00 \pm 1.25$ & $-2.30$ & \rule{-7pt}{0ex} & $12.50 \pm 1.12$ & $-1.89$ & \rule{-7pt}{0ex} & $12.68 \pm 0.60$ & $-1.97$ \\ 
   NGC5024 & $-8.71$ & $5.72$ & $12.67 \pm 0.64$ & $-1.86$ & \rule{-7pt}{0ex} & $13.25 \pm 0.50$ & $-2.00$ & \rule{-7pt}{0ex} & $12.25 \pm 1.03$ & $-2.06$ & \rule{-7pt}{0ex} & $12.72 \pm 0.44$ & $-1.97$ \\ 
   NGC5053 & $-6.76$ & $4.94$ & $12.29 \pm 0.51$ & $-1.98$ & \rule{-7pt}{0ex} & $13.50 \pm 0.75$ & $-2.40$ & \rule{-7pt}{0ex} & $12.25 \pm 1.07$ & $-2.30$ & \rule{-7pt}{0ex} & $12.68 \pm 0.47$ & $-2.23$ \\ 
   NGC5139 & $-10.26$ & $6.34$ & $11.52 \pm 0.64$ & $-1.35$ & \rule{-7pt}{0ex} & -- & -- & \rule{-7pt}{0ex} & -- & -- & \rule{-7pt}{0ex} & $11.52 \pm 0.64$ & $-1.35$ \\ 
   NGC5272 & $-8.88$ & $5.79$ & $11.39 \pm 0.51$ & $-1.34$ & \rule{-7pt}{0ex} & $12.50 \pm 0.50$ & $-1.60$ & \rule{-7pt}{0ex} & $11.75 \pm 1.03$ & $-1.50$ & \rule{-7pt}{0ex} & $11.88 \pm 0.42$ & $-1.48$ \\ 
   NGC5286 & $-8.74$ & $5.73$ & $12.54 \pm 0.51$ & $-1.41$ & \rule{-7pt}{0ex} & $13.00 \pm 1.00$ & $-1.70$ & \rule{-7pt}{0ex} & $12.50 \pm 1.07$ & $-1.70$ & \rule{-7pt}{0ex} & $12.68 \pm 0.52$ & $-1.60$ \\ 
   NGC5466 & $-6.98$ & $5.03$ & $13.57 \pm 0.64$ & $-2.20$ & \rule{-7pt}{0ex} & $13.00 \pm 0.75$ & $-2.10$ & \rule{-7pt}{0ex} & $12.50 \pm 1.03$ & $-2.31$ & \rule{-7pt}{0ex} & $13.02 \pm 0.48$ & $-2.20$ \\ 
   NGC5634 & $-7.69$ & $5.31$ & $11.84 \pm 0.51$ & $-1.94$ & \rule{-7pt}{0ex} & -- & -- & \rule{-7pt}{0ex} & -- & -- & \rule{-7pt}{0ex} & $11.84 \pm 0.51$ & $-1.94$ \\ 
   NGC5694 & $-7.83$ & $5.37$ & $13.44 \pm 0.90$ & $-1.74$ & \rule{-7pt}{0ex} & -- & -- & \rule{-7pt}{0ex} & -- & -- & \rule{-7pt}{0ex} & $13.44 \pm 0.90$ & $-1.74$ \\ 
   NGC5824 & $-8.85$ & $5.77$ & $12.80 \pm 0.90$ & $-1.60$ & \rule{-7pt}{0ex} & -- & -- & \rule{-7pt}{0ex} & -- & -- & \rule{-7pt}{0ex} & $12.80 \pm 0.90$ & $-1.60$ \\ 
   NGC5897 & $-7.23$ & $5.13$ & $12.30 \pm 1.20$ & $-1.73$ & \rule{-7pt}{0ex} & -- & -- & \rule{-7pt}{0ex} & -- & -- & \rule{-7pt}{0ex} & $12.30 \pm 1.20$ & $-1.73$ \\ 
   NGC5904 & $-8.81$ & $5.76$ & $10.62 \pm 0.38$ & $-1.12$ & \rule{-7pt}{0ex} & $12.25 \pm 0.75$ & $-1.30$ & \rule{-7pt}{0ex} & $11.50 \pm 1.03$ & $-1.33$ & \rule{-7pt}{0ex} & $11.46 \pm 0.44$ & $-1.25$ \\ 
   NGC5927 & $-7.81$ & $5.36$ & $12.67 \pm 0.90$ & $-0.64$ & \rule{-7pt}{0ex} & $12.25 \pm 0.75$ & $-0.50$ & \rule{-7pt}{0ex} & $10.75 \pm 1.07$ & $-0.29$ & \rule{-7pt}{0ex} & $11.89 \pm 0.53$ & $-0.48$ \\ 
   NGC5946 & $-7.18$ & $5.11$ & $11.39 \pm 0.90$ & $-1.22$ & \rule{-7pt}{0ex} & -- & -- & \rule{-7pt}{0ex} & -- & -- & \rule{-7pt}{0ex} & $11.39 \pm 0.90$ & $-1.22$ \\ 
   NGC5986 & $-8.44$ & $5.61$ & $12.16 \pm 0.51$ & $-1.35$ & \rule{-7pt}{0ex} & $13.25 \pm 1.00$ & $-1.60$ & \rule{-7pt}{0ex} & $12.25 \pm 1.25$ & $-1.63$ & \rule{-7pt}{0ex} & $12.55 \pm 0.56$ & $-1.53$ \\ 
   NGC6093 & $-8.23$ & $5.53$ & $12.54 \pm 0.51$ & $-1.47$ & \rule{-7pt}{0ex} & $13.50 \pm 1.00$ & $-1.70$ & \rule{-7pt}{0ex} & -- & -- & \rule{-7pt}{0ex} & $13.02 \pm 0.56$ & $-1.58$ \\ 
   NGC6101 & $-6.94$ & $5.01$ & $12.54 \pm 0.51$ & $-1.76$ & \rule{-7pt}{0ex} & $13.00 \pm 1.00$ & $-1.80$ & \rule{-7pt}{0ex} & $12.25 \pm 1.12$ & $-1.98$ & \rule{-7pt}{0ex} & $12.60 \pm 0.53$ & $-1.85$ \\ 
   NGC6121 & $-7.19$ & $5.11$ & $12.54 \pm 0.64$ & $-1.05$ & \rule{-7pt}{0ex} & $12.50 \pm 0.50$ & $-1.20$ & \rule{-7pt}{0ex} & $11.50 \pm 1.07$ & $-1.18$ & \rule{-7pt}{0ex} & $12.18 \pm 0.45$ & $-1.14$ \\ 
   NGC6144 & $-6.85$ & $4.97$ & $13.82 \pm 0.64$ & $-1.56$ & \rule{-7pt}{0ex} & $13.50 \pm 1.00$ & $-1.80$ & \rule{-7pt}{0ex} & $12.75 \pm 1.12$ & $-1.82$ & \rule{-7pt}{0ex} & $13.36 \pm 0.54$ & $-1.73$ \\ 
   NGC6171 & $-7.12$ & $5.08$ & $13.95 \pm 0.77$ & $-0.95$ & \rule{-7pt}{0ex} & $12.75 \pm 0.75$ & $-1.00$ & \rule{-7pt}{0ex} & $12.00 \pm 1.25$ & $-1.03$ & \rule{-7pt}{0ex} & $12.90 \pm 0.55$ & $-0.99$ \\ 
   NGC6205 & $-8.55$ & $5.65$ & $11.65 \pm 0.51$ & $-1.33$ & \rule{-7pt}{0ex} & $13.00 \pm 0.50$ & $-1.60$ & \rule{-7pt}{0ex} & $12.00 \pm 1.07$ & $-1.58$ & \rule{-7pt}{0ex} & $12.22 \pm 0.43$ & $-1.50$ \\ 
   NGC6218 & $-7.31$ & $5.16$ & $12.67 \pm 0.38$ & $-1.14$ & \rule{-7pt}{0ex} & $13.25 \pm 0.75$ & $-1.30$ & \rule{-7pt}{0ex} & $13.00 \pm 1.12$ & $-1.33$ & \rule{-7pt}{0ex} & $12.97 \pm 0.47$ & $-1.26$ \\ 
   NGC6235 & $-6.29$ & $4.75$ & $11.39 \pm 0.90$ & $-1.18$ & \rule{-7pt}{0ex} & -- & -- & \rule{-7pt}{0ex} & -- & -- & \rule{-7pt}{0ex} & $11.39 \pm 0.90$ & $-1.18$ \\ 
   NGC6254 & $-7.48$ & $5.23$ & $11.39 \pm 0.51$ & $-1.25$ & \rule{-7pt}{0ex} & $13.00 \pm 1.25$ & $-1.55$ & \rule{-7pt}{0ex} & $11.75 \pm 1.07$ & $-1.57$ & \rule{-7pt}{0ex} & $12.05 \pm 0.57$ & $-1.46$ \\ 
   NGC6266 & $-9.18$ & $5.91$ & $11.78 \pm 0.90$ & $-1.02$ & \rule{-7pt}{0ex} & -- & -- & \rule{-7pt}{0ex} & -- & -- & \rule{-7pt}{0ex} & $11.78 \pm 0.90$ & $-1.02$ \\ 
   NGC6273 & $-9.13$ & $5.89$ & $11.90 \pm 0.90$ & $-1.53$ & \rule{-7pt}{0ex} & -- & -- & \rule{-7pt}{0ex} & -- & -- & \rule{-7pt}{0ex} & $11.90 \pm 0.90$ & $-1.53$ \\ 
   NGC6284 & $-7.96$ & $5.42$ & $11.14 \pm 0.90$ & $-1.13$ & \rule{-7pt}{0ex} & -- & -- & \rule{-7pt}{0ex} & -- & -- & \rule{-7pt}{0ex} & $11.14 \pm 0.90$ & $-1.13$ \\ 
   NGC6287 & $-7.36$ & $5.18$ & $13.57 \pm 0.90$ & $-1.91$ & \rule{-7pt}{0ex} & -- & -- & \rule{-7pt}{0ex} & -- & -- & \rule{-7pt}{0ex} & $13.57 \pm 0.90$ & $-1.91$ \\ 
   NGC6304 & $-7.30$ & $5.15$ & $13.57 \pm 1.02$ & $-0.66$ & \rule{-7pt}{0ex} & $12.75 \pm 0.75$ & $-0.50$ & \rule{-7pt}{0ex} & $11.25 \pm 1.07$ & $-0.37$ & \rule{-7pt}{0ex} & $12.52 \pm 0.55$ & $-0.51$ \\ 
   NGC6341 & $-8.21$ & $5.52$ & $13.18 \pm 0.51$ & $-2.16$ & \rule{-7pt}{0ex} & $13.25 \pm 1.00$ & $-2.40$ & \rule{-7pt}{0ex} & $12.75 \pm 1.03$ & $-2.35$ & \rule{-7pt}{0ex} & $13.06 \pm 0.51$ & $-2.30$ \\ 
   NGC6342 & $-6.42$ & $4.80$ & $12.03 \pm 0.90$ & $-0.69$ & \rule{-7pt}{0ex} & -- & -- & \rule{-7pt}{0ex} & -- & -- & \rule{-7pt}{0ex} & $12.03 \pm 0.90$ & $-0.69$ \\ 
   NGC6352 & $-6.47$ & $4.82$ & $12.67 \pm 0.90$ & $-0.70$ & \rule{-7pt}{0ex} & $13.00 \pm 0.50$ & $-0.80$ & \rule{-7pt}{0ex} & $10.75 \pm 1.07$ & $-0.62$ & \rule{-7pt}{0ex} & $12.14 \pm 0.49$ & $-0.71$ \\ 
   NGC6362 & $-6.95$ & $5.01$ & $13.57 \pm 0.64$ & $-0.99$ & \rule{-7pt}{0ex} & $12.50 \pm 0.50$ & $-1.10$ & \rule{-7pt}{0ex} & $12.50 \pm 1.03$ & $-1.07$ & \rule{-7pt}{0ex} & $12.86 \pm 0.44$ & $-1.05$ \\ 
   NGC6366 & $-5.74$ & $4.53$ & $13.31 \pm 1.66$ & $-0.73$ & \rule{-7pt}{0ex} & $12.00 \pm 0.75$ & $-0.70$ & \rule{-7pt}{0ex} & $11.00 \pm 1.12$ & $-0.59$ & \rule{-7pt}{0ex} & $12.10 \pm 0.71$ & $-0.67$ \\ 
   NGC6388 & $-9.41$ & $6.00$ & $12.03 \pm 1.02$ & $-0.77$ & \rule{-7pt}{0ex} & -- & -- & \rule{-7pt}{0ex} & -- & -- & \rule{-7pt}{0ex} & $12.03 \pm 1.02$ & $-0.77$ \\ 
   NGC6397 & $-6.64$ & $4.89$ & $12.67 \pm 0.51$ & $-1.76$ & \rule{-7pt}{0ex} & $13.50 \pm 0.50$ & $-2.10$ & \rule{-7pt}{0ex} & $13.00 \pm 1.03$ & $-1.99$ & \rule{-7pt}{0ex} & $13.06 \pm 0.42$ & $-1.95$ \\ 
   NGC6426 & $-6.67$ & $4.90$ & $12.90 \pm 1.00$ & $-2.11$ & \rule{-7pt}{0ex} & $13.00 \pm 1.50$ & $-2.20$ & \rule{-7pt}{0ex} & -- & -- & \rule{-7pt}{0ex} & $12.95 \pm 0.90$ & $-2.16$ \\ 
   NGC6441 & $-9.63$ & $6.09$ & $11.26 \pm 0.90$ & $-0.60$ & \rule{-7pt}{0ex} & -- & -- & \rule{-7pt}{0ex} & -- & -- & \rule{-7pt}{0ex} & $11.26 \pm 0.90$ & $-0.60$ \\ 
   NGC6496 & $-7.20$ & $5.11$ & $12.42 \pm 0.90$ & $-0.70$ & \rule{-7pt}{0ex} & $12.00 \pm 0.75$ & $-0.50$ & \rule{-7pt}{0ex} & $10.75 \pm 1.07$ & $-0.46$ & \rule{-7pt}{0ex} & $11.72 \pm 0.53$ & $-0.55$ \\ 
   NGC6535 & $-4.75$ & $4.13$ & $10.50 \pm 1.15$ & $-1.51$ & \rule{-7pt}{0ex} & $13.25 \pm 1.00$ & $-1.90$ & \rule{-7pt}{0ex} & $12.75 \pm 1.12$ & $-1.79$ & \rule{-7pt}{0ex} & $12.17 \pm 0.63$ & $-1.73$ \\ 
   NGC6541 & $-8.52$ & $5.64$ & $12.93 \pm 0.51$ & $-1.53$ & \rule{-7pt}{0ex} & $13.25 \pm 1.00$ & $-1.90$ & \rule{-7pt}{0ex} & $12.50 \pm 1.12$ & $-1.82$ & \rule{-7pt}{0ex} & $12.89 \pm 0.53$ & $-1.75$ \\ 
   NGC6544 & $-6.94$ & $5.01$ & $10.37 \pm 0.90$ & $-1.20$ & \rule{-7pt}{0ex} & -- & -- & \rule{-7pt}{0ex} & -- & -- & \rule{-7pt}{0ex} & $10.37 \pm 0.90$ & $-1.20$ \\ 
   NGC6584 & $-7.69$ & $5.31$ & $11.26 \pm 0.38$ & $-1.30$ & \rule{-7pt}{0ex} & $12.25 \pm 0.75$ & $-1.40$ & \rule{-7pt}{0ex} & $11.75 \pm 1.12$ & $-1.50$ & \rule{-7pt}{0ex} & $11.75 \pm 0.47$ & $-1.40$ \\ 
   \hline
  \end{tabular} 
\end{table*}

\capstartfalse
\begin{table*}
  \contcaption{} 
\label{tab:gcobscont}
  \begin{tabular}{l c c c c c c c c c c c c c}
   \hline
 &  &  & \multicolumn{2}{c}{\citet{forbes10}} & \rule{-7pt}{0ex} & \multicolumn{2}{c}{\citet{dotter10,dotter11}} & \rule{-7pt}{0ex} & \multicolumn{2}{c}{\citet{vandenberg13}} & \rule{-7pt}{0ex} & \multicolumn{2}{c}{Mean} \vspace{1mm} \\ \cline{4-5} \cline{7-8} \cline{10-11} \cline{13-14} \rule{-2pt}{3ex} 
   Name & $M_V$ & $\log M$ & $\tau$ & $\feh$ & \rule{-7pt}{0ex} & $\tau$ & $\feh$ & \rule{-7pt}{0ex} & $\tau$ & $\feh$ & \rule{-7pt}{0ex} & $\overline{\tau}$ & $\overline{\feh}$ \\ 
    & $[{\rm mag}]$ & $[\msun]$ & $[\gyr]$ &  & \rule{-7pt}{0ex} & $[\gyr]$ &  & \rule{-7pt}{0ex} & $[\gyr]$ &  & \rule{-7pt}{0ex} & $[\gyr]$ &  \\ 
   \hline
   NGC6624 & $-7.49$ & $5.23$ & $12.54 \pm 0.90$ & $-0.70$ & \rule{-7pt}{0ex} & $13.00 \pm 0.75$ & $-0.50$ & \rule{-7pt}{0ex} & $11.25 \pm 1.12$ & $-0.42$ & \rule{-7pt}{0ex} & $12.26 \pm 0.54$ & $-0.54$ \\ 
   NGC6637 & $-7.64$ & $5.29$ & $13.06 \pm 0.90$ & $-0.78$ & \rule{-7pt}{0ex} & $12.50 \pm 0.75$ & $-0.70$ & \rule{-7pt}{0ex} & $11.00 \pm 1.07$ & $-0.59$ & \rule{-7pt}{0ex} & $12.19 \pm 0.53$ & $-0.69$ \\ 
   NGC6652 & $-6.66$ & $4.90$ & $12.93 \pm 0.77$ & $-0.97$ & \rule{-7pt}{0ex} & $13.25 \pm 0.50$ & $-0.75$ & \rule{-7pt}{0ex} & $11.25 \pm 1.03$ & $-0.76$ & \rule{-7pt}{0ex} & $12.48 \pm 0.46$ & $-0.83$ \\ 
   NGC6656 & $-8.50$ & $5.63$ & $12.67 \pm 0.64$ & $-1.49$ & \rule{-7pt}{0ex} & -- & -- & \rule{-7pt}{0ex} & -- & -- & \rule{-7pt}{0ex} & $12.67 \pm 0.64$ & $-1.49$ \\ 
   NGC6681 & $-7.12$ & $5.08$ & $12.80 \pm 0.51$ & $-1.35$ & \rule{-7pt}{0ex} & $13.00 \pm 0.75$ & $-1.50$ & \rule{-7pt}{0ex} & $12.50 \pm 1.12$ & $-1.70$ & \rule{-7pt}{0ex} & $12.77 \pm 0.48$ & $-1.52$ \\ 
   NGC6712 & $-7.50$ & $5.23$ & $10.40 \pm 1.40$ & $-0.94$ & \rule{-7pt}{0ex} & -- & -- & \rule{-7pt}{0ex} & -- & -- & \rule{-7pt}{0ex} & $10.40 \pm 1.40$ & $-0.94$ \\ 
   NGC6715 & $-9.98$ & $6.23$ & $10.75 \pm 0.38$ & $-1.25$ & \rule{-7pt}{0ex} & -- & -- & \rule{-7pt}{0ex} & $11.75 \pm 1.12$ & $-1.44$ & \rule{-7pt}{0ex} & $11.25 \pm 0.59$ & $-1.34$ \\ 
   NGC6717 & $-5.66$ & $4.50$ & $13.18 \pm 0.64$ & $-1.09$ & \rule{-7pt}{0ex} & $13.00 \pm 0.75$ & $-1.10$ & \rule{-7pt}{0ex} & $12.50 \pm 1.12$ & $-1.26$ & \rule{-7pt}{0ex} & $12.89 \pm 0.50$ & $-1.15$ \\ 
   NGC6723 & $-7.83$ & $5.37$ & $13.06 \pm 0.64$ & $-0.96$ & \rule{-7pt}{0ex} & $12.75 \pm 0.50$ & $-1.00$ & \rule{-7pt}{0ex} & $12.50 \pm 1.03$ & $-1.10$ & \rule{-7pt}{0ex} & $12.77 \pm 0.44$ & $-1.02$ \\ 
   NGC6752 & $-7.73$ & $5.33$ & $11.78 \pm 0.51$ & $-1.24$ & \rule{-7pt}{0ex} & $12.50 \pm 0.75$ & $-1.50$ & \rule{-7pt}{0ex} & $12.50 \pm 1.03$ & $-1.55$ & \rule{-7pt}{0ex} & $12.26 \pm 0.46$ & $-1.43$ \\ 
   NGC6779 & $-7.41$ & $5.20$ & $13.70 \pm 0.64$ & $-2.00$ & \rule{-7pt}{0ex} & $13.50 \pm 1.00$ & $-2.20$ & \rule{-7pt}{0ex} & $12.75 \pm 1.12$ & $-2.00$ & \rule{-7pt}{0ex} & $13.32 \pm 0.54$ & $-2.07$ \\ 
   NGC6809 & $-7.57$ & $5.26$ & $12.29 \pm 0.51$ & $-1.54$ & \rule{-7pt}{0ex} & $13.50 \pm 1.00$ & $-1.80$ & \rule{-7pt}{0ex} & $13.00 \pm 1.03$ & $-1.93$ & \rule{-7pt}{0ex} & $12.93 \pm 0.51$ & $-1.76$ \\ 
   NGC6838 & $-5.61$ & $4.48$ & $13.70 \pm 1.02$ & $-0.73$ & \rule{-7pt}{0ex} & $12.50 \pm 0.75$ & $-0.70$ & \rule{-7pt}{0ex} & $11.00 \pm 1.07$ & $-0.82$ & \rule{-7pt}{0ex} & $12.40 \pm 0.55$ & $-0.75$ \\ 
   NGC6864 & $-8.57$ & $5.66$ & $9.98 \pm 0.51$ & $-1.03$ & \rule{-7pt}{0ex} & -- & -- & \rule{-7pt}{0ex} & -- & -- & \rule{-7pt}{0ex} & $9.98 \pm 0.51$ & $-1.03$ \\ 
   NGC6934 & $-7.45$ & $5.21$ & $11.14 \pm 0.51$ & $-1.32$ & \rule{-7pt}{0ex} & $12.00 \pm 0.75$ & $-1.55$ & \rule{-7pt}{0ex} & $11.75 \pm 1.03$ & $-1.56$ & \rule{-7pt}{0ex} & $11.63 \pm 0.46$ & $-1.48$ \\ 
   NGC6981 & $-7.04$ & $5.05$ & $10.88 \pm 0.26$ & $-1.21$ & \rule{-7pt}{0ex} & $12.75 \pm 0.75$ & $-1.50$ & \rule{-7pt}{0ex} & $11.50 \pm 1.03$ & $-1.48$ & \rule{-7pt}{0ex} & $11.71 \pm 0.43$ & $-1.40$ \\ 
   NGC7006 & $-7.67$ & $5.30$ & -- & -- & \rule{-7pt}{0ex} & $12.25 \pm 0.75$ & $-1.50$ & \rule{-7pt}{0ex} & -- & -- & \rule{-7pt}{0ex} & $12.25 \pm 0.75$ & $-1.50$ \\ 
   NGC7078 & $-9.19$ & $5.91$ & $12.93 \pm 0.51$ & $-2.02$ & \rule{-7pt}{0ex} & $13.25 \pm 1.00$ & $-2.40$ & \rule{-7pt}{0ex} & $12.75 \pm 1.03$ & $-2.33$ & \rule{-7pt}{0ex} & $12.98 \pm 0.51$ & $-2.25$ \\ 
   NGC7089 & $-9.03$ & $5.85$ & $11.78 \pm 0.51$ & $-1.31$ & \rule{-7pt}{0ex} & $12.50 \pm 0.75$ & $-1.60$ & \rule{-7pt}{0ex} & $11.75 \pm 1.03$ & $-1.66$ & \rule{-7pt}{0ex} & $12.01 \pm 0.46$ & $-1.52$ \\ 
   NGC7099 & $-7.45$ & $5.21$ & $12.93 \pm 0.64$ & $-1.92$ & \rule{-7pt}{0ex} & $13.25 \pm 1.00$ & $-2.40$ & \rule{-7pt}{0ex} & $13.00 \pm 1.03$ & $-2.33$ & \rule{-7pt}{0ex} & $13.06 \pm 0.52$ & $-2.22$ \\ 
   NGC7492 & $-5.81$ & $4.56$ & $12.00 \pm 1.40$ & $-1.41$ & \rule{-7pt}{0ex} & -- & -- & \rule{-7pt}{0ex} & -- & -- & \rule{-7pt}{0ex} & $12.00 \pm 1.40$ & $-1.41$ \\ 
   Pal~1 & $-2.52$ & $3.24$ & $7.30 \pm 1.15$ & $-0.70$ & \rule{-7pt}{0ex} & -- & -- & \rule{-7pt}{0ex} & -- & -- & \rule{-7pt}{0ex} & $7.30 \pm 1.15$ & $-0.70$ \\ 
   Pal~3 & $-5.69$ & $4.51$ & $9.70 \pm 1.30$ & $-1.39$ & \rule{-7pt}{0ex} & $11.30 \pm 0.50$ & $-1.50$ & \rule{-7pt}{0ex} & -- & -- & \rule{-7pt}{0ex} & $10.50 \pm 0.70$ & $-1.44$ \\ 
   Pal~4 & $-6.01$ & $4.64$ & $9.50 \pm 1.60$ & $-1.07$ & \rule{-7pt}{0ex} & $10.90 \pm 0.50$ & $-1.30$ & \rule{-7pt}{0ex} & -- & -- & \rule{-7pt}{0ex} & $10.20 \pm 0.84$ & $-1.18$ \\ 
   Pal~5 & $-5.17$ & $4.30$ & $9.80 \pm 1.40$ & $-1.24$ & \rule{-7pt}{0ex} & $12.00 \pm 1.00$ & $-1.40$ & \rule{-7pt}{0ex} & -- & -- & \rule{-7pt}{0ex} & $10.90 \pm 0.86$ & $-1.32$ \\ 
   Pal~12 & $-4.47$ & $4.02$ & $8.83 \pm 1.12$ & $-0.83$ & \rule{-7pt}{0ex} & $9.50 \pm 0.75$ & $-0.80$ & \rule{-7pt}{0ex} & $9.00 \pm 1.07$ & $-0.81$ & \rule{-7pt}{0ex} & $9.11 \pm 0.57$ & $-0.81$ \\ 
   Pal~14 & $-4.80$ & $4.15$ & $10.50 \pm 1.00$ & $-1.36$ & \rule{-7pt}{0ex} & $10.50 \pm 0.50$ & $-1.50$ & \rule{-7pt}{0ex} & -- & -- & \rule{-7pt}{0ex} & $10.50 \pm 0.56$ & $-1.43$ \\ 
   Pal~15 & $-5.51$ & $4.44$ & -- & -- & \rule{-7pt}{0ex} & $13.00 \pm 1.50$ & $-2.00$ & \rule{-7pt}{0ex} & -- & -- & \rule{-7pt}{0ex} & $13.00 \pm 1.50$ & $-2.00$ \\ 
   Terzan~7 & $-5.01$ & $4.24$ & $7.30 \pm 0.51$ & $-0.56$ & \rule{-7pt}{0ex} & $8.00 \pm 0.75$ & $-0.60$ & \rule{-7pt}{0ex} & -- & -- & \rule{-7pt}{0ex} & $7.65 \pm 0.45$ & $-0.58$ \\ 
   Terzan~8 & $-5.07$ & $4.26$ & $12.16 \pm 0.51$ & $-1.80$ & \rule{-7pt}{0ex} & $13.50 \pm 0.50$ & $-2.40$ & \rule{-7pt}{0ex} & $13.00 \pm 1.07$ & $-2.34$ & \rule{-7pt}{0ex} & $12.89 \pm 0.43$ & $-2.18$ \\ 
   AM~1 & $-4.73$ & $4.13$ & $11.10 \pm 1.00$ & $-1.47$ & \rule{-7pt}{0ex} & $11.10 \pm 0.50$ & $-1.50$ & \rule{-7pt}{0ex} & -- & -- & \rule{-7pt}{0ex} & $11.10 \pm 0.56$ & $-1.48$ \\ 
   AM~4 & $-1.81$ & $2.96$ & $9.00 \pm 0.50$ & $-0.97$ & \rule{-7pt}{0ex} & -- & -- & \rule{-7pt}{0ex} & -- & -- & \rule{-7pt}{0ex} & $9.00 \pm 0.50$ & $-0.97$ \\ 
   Arp~2 & $-5.29$ & $4.35$ & $10.88 \pm 0.77$ & $-1.45$ & \rule{-7pt}{0ex} & $13.00 \pm 0.75$ & $-1.80$ & \rule{-7pt}{0ex} & $12.00 \pm 1.07$ & $-1.74$ & \rule{-7pt}{0ex} & $11.96 \pm 0.51$ & $-1.66$ \\ 
   E~3 & $-4.12$ & $3.88$ & $12.80 \pm 1.41$ & $-0.83$ & \rule{-7pt}{0ex} & -- & -- & \rule{-7pt}{0ex} & -- & -- & \rule{-7pt}{0ex} & $12.80 \pm 1.41$ & $-0.83$ \\ 
   Eridanus & $-5.13$ & $4.29$ & $8.90 \pm 1.60$ & $-1.20$ & \rule{-7pt}{0ex} & $10.50 \pm 0.50$ & $-1.30$ & \rule{-7pt}{0ex} & -- & -- & \rule{-7pt}{0ex} & $9.70 \pm 0.84$ & $-1.25$ \\ 
   IC4499 & $-7.32$ & $5.16$ & -- & -- & \rule{-7pt}{0ex} & $12.00 \pm 0.75$ & $-1.60$ & \rule{-7pt}{0ex} & -- & -- & \rule{-7pt}{0ex} & $12.00 \pm 0.75$ & $-1.60$ \\ 
   Lynga~7 & $-6.60$ & $4.87$ & $14.46 \pm 1.79$ & $-0.64$ & \rule{-7pt}{0ex} & $12.50 \pm 1.00$ & $-0.60$ & \rule{-7pt}{0ex} & -- & -- & \rule{-7pt}{0ex} & $13.48 \pm 1.03$ & $-0.62$ \\ 
   Pyxis & $-5.73$ & $4.53$ & -- & -- & \rule{-7pt}{0ex} & $11.50 \pm 1.00$ & $-1.50$ & \rule{-7pt}{0ex} & -- & -- & \rule{-7pt}{0ex} & $11.50 \pm 1.00$ & $-1.50$ \\ 
   Rup~106 & $-6.35$ & $4.77$ & $10.20 \pm 1.40$ & $-1.49$ & \rule{-7pt}{0ex} & $11.50 \pm 0.50$ & $-1.50$ & \rule{-7pt}{0ex} & -- & -- & \rule{-7pt}{0ex} & $10.85 \pm 0.74$ & $-1.50$ \\ 
   Whiting~1 & $-2.46$ & $3.22$ & $6.50 \pm 0.75$ & $-0.65$ & \rule{-7pt}{0ex} & -- & -- & \rule{-7pt}{0ex} & -- & -- & \rule{-7pt}{0ex} & $6.50 \pm 0.75$ & $-0.65$ \\ 
   \hline
  \end{tabular} 
\end{table*}
\capstarttrue

\bsp

\label{lastpage}

\end{document}